\documentclass[12pt]{article}

\font\grande=cmr9.5 scaled \magstep4
\font\medio=cmr9.5 scaled \magstep2
\outer\def\beginsection#1\par{\medbreak\bigskip
      \message{#1}\leftline{\bf#1}\nobreak\medskip
\vskip-\parskip
      \noindent}
\usepackage{graphicx} 
\begin{document}
\bibliographystyle {unsrt}

\titlepage

\begin{flushright}
CERN-PH-TH/2011-067
\end{flushright}

\vspace{10mm}
\begin{center}
{\grande Primordial vorticity and gradient expansion}\\
\vspace{1.5cm}
 Massimo Giovannini$^{a,b}$\footnote{Electronic address: massimo.giovannini@cern.ch} 
 and Zahra Rezaei$^{a,c}$\footnote{Electronic address: zahra.rezaei@cern.ch}\\
\vspace{1cm}
{{\sl $^{a}$Department of Physics, 
Theory Division, CERN, 1211 Geneva 23, Switzerland }}\\
\vspace{0.5cm}
{{\sl $^b$ INFN, Section of Milan-Bicocca, 20126 Milan, Italy}}\\
\vspace{0.5cm}
{{\sl $^c$ Isfahan University of Technology Department of Physics, 84154 Isfahan, Iran }}
\vspace*{0.5cm}
\end{center}

\vskip 0.5cm
\centerline{\medio  Abstract}
The evolution equations of the vorticities of the electrons, ions and photons in a pre-decoupling plasma are derived, in a fully inhomogeneous geometry,  by combining the general relativistic gradient expansion and the drift approximation
within the Adler-Misner-Deser decomposition.
The vorticity transfer between the different species is discussed 
in this novel framework and a set of general conservation laws, 
connecting the vorticities of the three-component plasma with the magnetic field intensity, is derived.
After demonstrating that a source of large-scale vorticity resides in the spatial gradients of the geometry and of the electromagnetic sources, the total vorticity is estimated to lowest order in the spatial gradients and by enforcing the validity of the momentum constraint. By acknowledging the current bounds 
on the tensor to scalar ratio in the (minimal) tensor extension of the $\Lambda$CDM paradigm the 
maximal comoving magnetic field induced by the total vorticity turns out to be, at most, of the order 
of $10^{-37}$ G over the typical comoving scales ranging between $1$ and 
$10$ Mpc. While the obtained results seem to be irrelevant for seeding a reasonable
galactic dynamo action, they demonstrate how the proposed fully inhomogeneous 
treatment can be used for the systematic scrutiny of pre-decoupling plasmas beyond 
the conventional perturbative expansions.
\noindent

\vspace{5mm}

\vfill
\newpage
\renewcommand{\theequation}{1.\arabic{equation}}
\setcounter{equation}{0}
\section{Primordial vorticities?}
\label{sec1}
The observed Universe might originate from a strongly coupled electromagnetic plasma 
existing prior to photon decoupling where the angular momentum transfer between 
ions, electrons and photons in an expanding space-time geometry  leads
to the formation of large-scale vortices as speculated by 
various authors including, with slightly different perspectives, 
Hoyle \cite{hoyle},  Harrison \cite{harrison1,harrison2,harrison3}, Mishustin and Ruzmaikin \cite{mishustin}, Ozernoy and Chernin \cite{ozernoy1,ozernoy2,ozernoy3} 
and others (see also \cite{peebles}).  The primordial vorticity, if present in the 
pre-decoupling plasma, might lead eventually to the 
formation of large-scale magnetic fields possibly relevant for galactic magnetogenesis. 

The physical description of the angular momentum 
exchange between ions, electrons and photons can be realized by
appropriately translating the evolution equations describing ionized 
gases \cite{spitzer,krall} to an expanding geometry supplemented 
by its own relativistic fluctuations \cite{mg1,mg1aa}. In the context of the $\Lambda$CDM 
paradigm\footnote{The acronym $\Lambda$CDM (where 
$\Lambda$ denotes the dark energy component and CDM stands for cold dark matter) and the 
terminology concordance paradigm will be used interchangeably.}, it  
is both reasonable and justified to assume that the  background 
geometry is conformally flat and that its inhomogeneities stem from the relativistic 
fluctuations of the spatial curvature described  either in gauge-invariant terms or in an appropriate 
gauge. The latter assumption rests 
exactly on the absence of large-scale vorticity which is assumed to be vanishing at least 
within the current observational precision. 

A gross argument could suggest that the vorticity must be negligible for $\Lambda$CDM initial conditions,  since it is 
the curl of a velocity. Thanks to the momentum constraint (connecting the first derivatives of the linearized fluctuations of the geometry to the peculiar velocities), 
the total velocity field is subleading when compared with the density contrasts or with the curvature perturbation for typical scales larger than the Hubble radius and in the case of the conventional adiabatic initial conditions postulated in the vanilla $\Lambda$CDM scenario.  The latter argument suggests that the treatment 
of large-scale vorticity assumes, more or less tacitly, a correct treatment of the spatial gradients.
To transform this incomplete observation in a more rigorous approach
it is necessary to introduce a description of the vorticity which does not rely on the 
purported smallness or largeness of the gravitational fluctuations. It is 
rather desirable to describe the angular momentum exchange 
between ions, electrons and photons in a gravitating plasma which is also fully inhomogeneous.  By 
fully inhomogeneous plasma we mean the situation where 
not only the concentrations of charged and neutral species 
depend, in an arbitrary manner, upon the spatial coordinates
but where the geometry as well as the electromagnetic 
fields are not homogeneous. It has been recently 
argued \cite{mg2} that such a description can be rather effective 
for the analysis of a wide range of phenomena including the 
physics of pre-decoupling plasmas. In the present 
paper  the results of Ref. \cite{mg2} shall be first extended and then applied to a 
concrete situation with the 
purpose of obtaining  an explicit set of equations describing the evolution of the vorticities of the various species of the plasma. 
The proposal of \cite{mg2} is built on the fully inhomogeneous description 
of the geometry in terms of the Adler-Misner-Deser (ADM) variables \cite{ADM1,ADM2} which are customarily exploited for the implementation of the general relativistic gradient expansion \cite{gr1,gr2,gr3,gr4,gr5,gr6}.
The second key ingredient of Ref. \cite{mg2} is the fully inhomogeneous description of cold plasmas in flat space which is the starting points of the analysis of nonlinear effects in kinetic theory and in magnetohydrodynamics 
(see, e.g. \cite{gr7,gr8,gr9}).  
Consequently, the vorticity exchange 
between ions, electrons and photons can be analyzed in gravitating plasmas
with the help of an expansion scheme which involves not only the gradients of the geometry, but also 
the gradients of the electromagnetic sources, by so combining the general relativistic gradient expansion and the drift approximation (sometimes dubbed giuding center approximation) typical of cold plasmas.

The approach pursued in this paper reproduces, in the conformally flat limit,
the conventional treatment which will be made more precise in section \ref{sec2}.  
The evolution of the vorticity in gravitating plasmas which are also fully inhomogeneous will be 
discussed in section \ref{sec3}.   In sections \ref{sec4}  and \ref{sec5}  the total vorticity 
of the geometry will be computed within the gradient expansion and estimated in the 
framework of the $\Lambda$CDM paradigm. The maximal magnetic field 
induced by the total vorticity will be computed in section \ref{sec6}. 
Section \ref{sec7} contains our concluding remarks. In appendix \ref{APPA} 
some useful complements have been included to make the paper self-contained while 
in appendix \ref{APPB} useful details on the calculations of correlation 
functions of multiple fields in real space have been included for the technical 
benefit of the interested readers. 

\renewcommand{\theequation}{2.\arabic{equation}}
\setcounter{equation}{0}
\section{Vorticities in conventional perturbative expansions}
\label{sec2}
The treatment proposed here differs slightly from the one of 
 Refs. \cite{harrison1,harrison2,harrison3,mishustin} for three reasons:
(i) the conformal time coordinate is preferred to  the cosmic time;
(ii) the relativistic fluctuations of the geometry are included in the longitudinal gauge;
(iii) the three-fluid, two-fluids and one fluid descriptions are 
discussed more explicitly within the appropriate temperature ranges where 
they are applicable. 

The conformal
flatness of the geometry does not imply the invariance 
of the system under the Weyl rescaling of the metric. Such a potential symmetry is broken 
by the masses of the electrons and ions which are crucial in the large-scale 
evolution of the vorticity. The considerations of the present section can also be formulated in the case of a geometry which is not spatially flat; this is not essential  since the subsequent generalizations will automatically 
include also geometries which are not necessarily spatially flat. 
Consider first the case of a conformally flat background geometry characterized 
by a metric tensor $\overline{g}_{\mu\nu}= a^2(\tau)\eta_{\mu\nu}$ and supplemented 
by the corresponding relativistic fluctuations which we write in the longitudinal gauge 
\begin{equation}
\delta_{\mathrm{s}} g_{00}(\vec{x},\tau) = 2\, a^2(\tau)  \phi(\vec{x},\tau),\qquad \delta_{\mathrm{s}} g_{ij}(\vec{x},\tau) = 2\,a^2(\tau) 
\psi(\vec{x},\tau) \delta_{ij};
\label{met1}
\end{equation}
note that $\delta_{\mathrm{s}}$ describes a metric perturbation which preserves the scalar nature of the 
fluctuation since, in the $\Lambda$CDM paradigm, the dominant source of inhomogeneity 
comes from the scalar modes of the geometry. 
By defining the comoving electromagnetic fields $\vec{E}$ and $\vec{B}$ as well
as the comoving concentrations of electrons and ions (i.e. $n_{\mathrm{e}}$ and $n_{\mathrm{i}}$)
\begin{eqnarray}
&& \vec{E}(\vec{x},\tau) = a^2(\tau) \vec{{\mathcal E}}(\vec{x},\tau), \qquad  \vec{B}(\vec{x},\tau) = a^2(\tau) \vec{{\mathcal B}}(\vec{x},\tau),
\nonumber\\
&& n_{\mathrm{i}}(\vec{x},\tau) = a^3(\tau) \tilde{n}_{\mathrm{i}}(\vec{x},\tau),\qquad 
n_{\mathrm{e}}(\vec{x},\tau) = a^3(\tau) \tilde{n}_{\mathrm{e}}(\vec{x},\tau),
\label{S1}
\end{eqnarray}
Maxwell's equations read
\begin{eqnarray}
&& \vec{\nabla} \cdot \vec{E} = 4 \pi e (n_{\mathrm{i}} - n_{\mathrm{e}}),\qquad\vec{\nabla} \cdot \vec{B} =0,
\label{S2}\\
&& \vec{\nabla} \times \vec{E} = - \partial_{\tau} \vec{B}, \qquad  \vec{\nabla}\times \vec{B} = 4\pi e (n_{\mathrm{i}}\, \vec{v}_{\mathrm{i}} - 
n_{\mathrm{e}}\, \vec{v}_{\mathrm{e}} ) + \partial_{\tau}\vec{E}.
\label{S3}
\end{eqnarray}
In Eqs. (\ref{S1}), (\ref{S2}) and (\ref{S3}) all the fields are appropriately rescaled so that 
the resulting equations are formally equivalent to the ones of the flat space-time.  The peculiar velocities of the ions, electrons and photons obey the following set of 
equations\footnote{As usual ${\mathcal H} = \partial_{\tau} \ln{a}$ and its relation 
with the Hubble rate is simply $ {\mathcal H} = a H$.} 
\begin{eqnarray}
&& \partial_{\tau}\vec{v}_{\mathrm{e}} + {\mathcal H}\,\vec{v}_{\mathrm{e}} = - \frac{e n_{\mathrm{e}}}{\rho_{\mathrm{e}} \, a^{4}} [ \vec{E} + \vec{v}_{\mathrm{e}} \times \vec{B}] - \vec{\nabla} \phi 
+ 
\frac{4}{3} \frac{\rho_{\gamma}}{\rho_{\mathrm{e}}} a 
\Gamma_{\gamma \, \mathrm{e}} (\vec{v}_{\gamma} - \vec{v}_{\mathrm{e}}) + a \Gamma_{\mathrm{e\,i}} ( \vec{v}_{\mathrm{i}} - \vec{v}_{\mathrm{e}}),
\label{SA}\\
&&  \partial_{\tau} \vec{v}_{\mathrm{i}} + {\mathcal H}\,\vec{v}_{\mathrm{i}} =   \frac{e n_{\mathrm{i}}}{\rho_{\mathrm{i}} \, a^{4}}[ \vec{E} + \vec{v}_{\mathrm{i}} \times \vec{B}] - \vec{\nabla} \phi 
+ 
\frac{4}{3} \frac{\rho_{\gamma}}{\rho_{\mathrm{i}}} a 
\Gamma_{\gamma \, \mathrm{i}} (\vec{v}_{\gamma}-\vec{v}_{\mathrm{i}} ) + a \Gamma_{\mathrm{e\,i}} \frac{\rho_{\mathrm{e}}}{\rho_{\mathrm{i}}}( \vec{v}_{\mathrm{e}} - \vec{v}_{\mathrm{i}}),
\label{SB}\\
&& \partial_{\tau} \vec{v}_{\gamma} = - \frac{1}{4} \vec{\nabla} \delta_{\gamma} - \vec{\nabla} \phi 
+ a \Gamma_{\gamma\mathrm{i}} (\vec{v}_{\mathrm{i}} - \vec{v}_{\gamma}) + 
a \Gamma_{\gamma\mathrm{e}}  ( \vec{v}_{\mathrm{e}} - \vec{v}_{\gamma}).
\label{SC}
\end{eqnarray}
In Eqs. (\ref{SA})--(\ref{SC}) the relativistic fluctuations of the geometry are included from the very beginning in terms of the longitudinal gauge variables of Eq. (\ref{met1}); the electron-photon,  electron-ion and ion-photon 
rates of momentum exchange appearing in Eqs. (\ref{SA})--(\ref{SC}) are given by\footnote{Note that $T$ denotes the temperature and $\Lambda_{\mathrm{C}}$ is the Coulomb logarithm \cite{spitzer,krall}.}:
\begin{eqnarray}
&&\Gamma_{\gamma\mathrm{e}} = \tilde{n}_{\mathrm{e}} 
\sigma_{\mathrm{e}\gamma},\qquad 
\Gamma_{\gamma\mathrm{i}} = \tilde{n}_{\mathrm{i}} 
\sigma_{\mathrm{i}\gamma},\qquad \sigma_{\mathrm{e}\gamma} 
= \frac{8}{3}\pi \biggl(\frac{e^2}{m_{\mathrm{e}}}\biggr)^2, \qquad 
\sigma_{\mathrm{i}\gamma} 
= \frac{8}{3}\pi \biggl(\frac{e^2}{m_{\mathrm{i}}}\biggr)^2,
\label{S9}\\
&& \Gamma_{\mathrm{e\,i}} = \tilde{n}_{\mathrm{e}} \sqrt{\frac{T}{m_{\mathrm{e}}}} \, \sigma_{\mathrm{e\,i}} = \Gamma_{\mathrm{i\, e}},\qquad \sigma_{\mathrm{e\,i}} = 
\frac{e^4}{T^2} \ln{\Lambda_{\mathrm{C}}},\qquad \Lambda_{\mathrm{C}} = \frac{3}{2 e^3} \sqrt{\frac{T^3}{\tilde{n}_{\mathrm{e}}\pi}}.
\label{S10}
\end{eqnarray}
Note that, in Eq. (\ref{S9}) and (\ref{S10}), $T$ and $\tilde{n}$ are, respectively,  
physical temperatures and physical concentrations. If the rates and the cross sections 
would be consistently expressed in terms of comoving temperatures $\overline{T} = a T$ and 
comoving concentrations $n = a^3 \, \tilde{n}$ the corresponding rates will inherit a scale factor for each 
mass. For instance $a\Gamma_{\mathrm{e\,i}}$ becomes $n_{\mathrm{e}} \, \sqrt{\overline{T}/(m_{\mathrm{e}} a)} \, (e^4/\overline{T}^2) \, \ln{\Lambda_{\mathrm{C}}}$,  if comoving temperature and concentrations are used. 

Let us then define the vorticities associated with the peculiar velocities of the various species
\begin{equation}
\vec{\omega}_{\mathrm{e}}(\vec{x},\tau) = \vec{\nabla} \times \vec{v}_{\mathrm{e}},
 \qquad \vec{\omega}_{\mathrm{i}}(\vec{x},\tau)= \vec{\nabla} \times \vec{v}_{\mathrm{i}}, \qquad 
\vec{\omega}_{\gamma}(\vec{x},\tau) = \vec{\nabla} \times \vec{v}_{\gamma},
\label{S4}
\end{equation}
and their corresponding three-divergences:
\begin{equation}
\theta_{\mathrm{e}}(\vec{x},\tau) = \vec{\nabla} \cdot \vec{v}_{\mathrm{e}}, \qquad \theta_{\mathrm{i}}(\vec{x},\tau) = \vec{\nabla}\cdot \vec{v}_{\mathrm{i}}, \qquad 
\theta_{\gamma}(\vec{x},\tau) = \vec{\nabla} \cdot \vec{v}_{\gamma}.
\label{S5}
\end{equation}
The evolution equations of the vorticities and of the divergences can be obtained by taking, respectively, the 
curl and the divergence of Eqs. (\ref{SA})--(\ref{SC}) and by using Eqs. (\ref{S2}) and (\ref{S3}). To simplify 
the obtained expressions it is useful to introduce the total comoving charge density and the comoving current density. 
\begin{equation}
\rho_{\mathrm{q}}= e(n_{\mathrm{i}} - n_{\mathrm{e}}),\qquad 
\vec{J} = e ( n_{\mathrm{i}} \vec{v}_{\mathrm{i}} - n_{\mathrm{e}} \vec{v}_{\mathrm{e}}).
\label{th1a}
\end{equation}
Thus, the evolution of the vorticities and of the divergences of the electrons are, respectively,
\begin{eqnarray}
\partial_{\tau}\vec{\omega}_{\mathrm{e}}+ {\mathcal H}\,\vec{\omega}_{\mathrm{e}} &=& \frac{e n_{\mathrm{e}}}{\rho_{\mathrm{e}} \, a^{4}} \biggl[ \partial_{\tau} \vec{B} 
+ (\vec{v}_{\mathrm{e}}\cdot\vec{\nabla}) \vec{B} + \theta_{\mathrm{e}} \vec{B} - (\vec{B} \cdot\vec{\nabla})\vec{v}_{\mathrm{e}}\biggr]  
\nonumber\\
&+& 
\frac{4}{3} \frac{\rho_{\gamma}}{\rho_{\mathrm{e}}} a 
\Gamma_{\gamma \, \mathrm{e}} (\vec{\omega}_{\gamma} - \vec{\omega}_{\mathrm{e}}) + a \Gamma_{\mathrm{e\,i}} ( \vec{\omega}_{\mathrm{i}} - \vec{\omega}_{\mathrm{e}}),
\label{S6}\\
\partial_{\tau} \theta_{\mathrm{e}} + {\mathcal H} \theta_{\mathrm{e}} &=& - \frac{e n_{\mathrm{e}}}{\rho_{\mathrm{e}} \, a^{4}} \biggl[ 4 \pi \rho_{\mathrm{q}} + 
\vec{\omega}_{\mathrm{e}} \cdot \vec{B} - 4 \pi \vec{v}_{\mathrm{e}} \cdot \vec{J} - \vec{v}_{\mathrm{e}} \cdot \partial_{\tau} \vec{E}\biggr] - \nabla^2 \phi  
\nonumber\\
&+& \frac{4}{3} \frac{\rho_{\gamma}}{\rho_{\mathrm{e}}} a \Gamma_{\gamma\mathrm{e}} (\theta_{\gamma} - \theta_{\mathrm{e}})+ 
a \Gamma_{\mathrm{e}\,\mathrm{i}} (\theta_{\mathrm{i}} - \theta_{\mathrm{e}}),
\label{th1}
\end{eqnarray}
Conversely the vorticity and the three-divergence of the ions evolve as:
\begin{eqnarray}
\partial_{\tau} \vec{\omega}_{\mathrm{i}} + {\mathcal H}\,\vec{\omega}_{\mathrm{i}} &=&  - \frac{e n_{\mathrm{i}}}{\rho_{\mathrm{i}} \, a^{4}} 
\biggl[ \partial_{\tau} \vec{B} + (\vec{v}_{\mathrm{i}}\cdot\vec{\nabla}) \vec{B} + \theta_{\mathrm{i}} \vec{B} - (\vec{B} \cdot\vec{\nabla})\vec{v}_{\mathrm{i}}\biggr]
\nonumber\\
&+& \frac{4}{3} \frac{\rho_{\gamma}}{\rho_{\mathrm{i}}} a 
\Gamma_{\gamma \, \mathrm{i}} (\vec{\omega}_{\gamma}-\vec{\omega}_{\mathrm{i}} ) 
+ a \Gamma_{\mathrm{e\,i}} \frac{\rho_{\mathrm{e}}}{\rho_{\mathrm{i}}}( \vec{\omega}_{\mathrm{e}} - \vec{\omega}_{\mathrm{i}}),
\label{S7}\\
\partial_{\tau} \theta_{\mathrm{i}} + {\mathcal H} \theta_{\mathrm{i}} &=& \frac{e n_{\mathrm{i}}}{\rho_{\mathrm{i}} \, a^{4}} \biggl[ 4 \pi \rho_{\mathrm{q}} + 
\vec{\omega}_{\mathrm{i}} \cdot \vec{B} - 4 \pi \vec{v}_{\mathrm{i}} \cdot \vec{J} - \vec{v}_{\mathrm{i}} \cdot \partial_{\tau} \vec{E}\biggr] - \nabla^2 \phi
\nonumber\\
&+& \frac{4}{3} \frac{\rho_{\gamma}}{\rho_{\mathrm{i}}} a \Gamma_{\gamma\mathrm{i}} (\theta_{\gamma} - \theta_{\mathrm{i}})+ 
a \Gamma_{\mathrm{e}\,\mathrm{i}} \frac{\rho_{\mathrm{e}}}{\rho_{\mathrm{i}}} (\theta_{\mathrm{e}} - \theta_{\mathrm{i}}).
\label{th2}
\end{eqnarray}
Finally, the evolution equations for the photons are given by:
\begin{eqnarray}
\partial_{\tau}\vec{\omega}_{\gamma} &=&  a \Gamma_{\gamma\mathrm{i}} (\vec{\omega}_{\mathrm{i}} - \vec{\omega}_{\gamma}) + 
a \Gamma_{\gamma\mathrm{e}}  ( \vec{\omega}_{\mathrm{e}} - \vec{\omega}_{\gamma}),
\label{S8}\\
\partial_{\tau} \theta_{\gamma} &=& - \frac{1}{4}\nabla^2 \delta_{\gamma} - \nabla^2 \phi  + 
a \Gamma_{\gamma\mathrm{i}} (\theta_{\mathrm{i}} - \theta_{\gamma}) + 
a \Gamma_{\gamma\mathrm{e}}  ( \theta_{\mathrm{e}} - \theta_{\mathrm{i}}).
\label{th3}
\end{eqnarray}
The system described by the set of equations deduced so far will be considered as globally neutral. In particular, prior to photon decoupling,
the electron and ion (comoving) concentrations have a common value $n_{0}$, i.e. $n_{\mathrm{i}} = n_{\mathrm{e}} = n_{0}$ where\footnote{If not otherwise stated the pivotal values of the cosmological parameters will be the ones determined 
from the WMAP 7yr data alone in the light of the $\Lambda$CDM paradigm.}
\begin{equation}
 n_{0}= \eta_{\mathrm{b}0} n_{\gamma}, \qquad \eta_{\mathrm{b}0}=6.177 \times 10^{-10} 
 \biggl(\frac{h_{0}^2\Omega_{\mathrm{b}0}}{0.02258}\biggr) \biggl(\frac{2.725\, \mathrm{K}}{T_{\gamma 0}}
 \biggr)^{3},
\label{th4}
\end{equation} 
and $T_{\gamma 0}$ is the present value of the CMB temperature determining the concentration of the photons; $\Omega_{\mathrm{b}0}$ is the present value of the critical fraction of baryons, while $h_{0}$ is the Hubble constant in units 
of $100\, \mathrm{Km}/(\mathrm{Mpc} \times \mathrm{sec})$.  
The system of Eqs. (\ref{S6})--(\ref{th3}) is coupled 
with the evolution of the density contrasts of the electrons, ions and photons (i.e. 
$\delta_{\mathrm{e}}$, $\delta_{\mathrm{i}}$ and $\delta_{\gamma}$) 
\begin{eqnarray}
&& \delta_{\mathrm{e}}' = - \theta_{\mathrm{e}} + 3 \psi' 
- \frac{e n_{\mathrm{e}}}{\rho_{\mathrm{e}} a^4}\vec{E} \cdot \vec{v}_{\mathrm{e}},\qquad  \delta_{\mathrm{i}}' = - \theta_{\mathrm{i}} + 3 \psi' 
+ \frac{e n_{\mathrm{i}}}{\rho_{\mathrm{i}} a^4}\vec{E} \cdot \vec{v}_{\mathrm{i}},
\label{dc1}\\
&& \delta_{\gamma}' = 4\psi' - \frac{4}{3} \theta_{\gamma}.
\label{dc2}
\end{eqnarray}
Finally the metric fluctuations, the density contrasts and the divergences of the peculiar 
velocities are both determined and constrained by the perturbed Einstein equations 
(see, e.g. Eqs. (2.43)--(2.46) in the first article of Ref. \cite{mg1}). 
Concerning the system of Eqs. (\ref{S6})--(\ref{th3})  two comments are in order:
\begin{itemize}
\item{} Eqs. (\ref{S6})--(\ref{th1}) (as well as Eqs. (\ref{S7})--(\ref{th2})) couple together the evolution of the vorticities, the evolution of the
divergences and the gradients of the magnetic field; while in the linearized approximation 
the spatial gradients are simply neglected, in the forthcoming sections the 
evolution of the vorticity will be studied
to a given order in the spatial gradients;
\item{} the electron and ion masses break the Weyl rescaling of the whole system 
of equations; this aspect can be appreciated by noticing that the prefactor appearing in front of the square brackets at the right hand side of Eqs. (\ref{S6})--(\ref{th1}) 
and Eqs. (\ref{S7})--(\ref{th2})
is, respectively,  $e/(m_{\mathrm{e}} a)$ and $e/(m_{\mathrm{i}} a)$.
\end{itemize}
Equations (\ref{S6})--(\ref{th3}) have three different 
scales of vorticity exchange: the photon-ion, the photon-electron and the electron ion rates whose 
respective magnitude determines the subleading terms and the different dynamical regimes. By taking the ratios of the two rates appearing at the right hand side of Eqs. (\ref{S6}) and 
 (\ref{S7}) the following two dimensionless ratios can be constructed\footnote{Note that $\rho_{\mathrm{i}}$ must 
 simplify when taking the ratio of the two rates in Eq. (\ref{S7}).}:
\begin{eqnarray}
\frac{3 \rho_{\mathrm{e}} \, \Gamma_{\mathrm{e\, i}}}{4 \, \rho_{\gamma} \Gamma_{\gamma\mathrm{e}}} &=& \frac{135 \, \zeta(3)}{16 \, \pi^5} \biggl(\frac{T}{m_{\mathrm{e}}}\biggr)^{-5/2}\, \eta_{\mathrm{b} 0} \, \ln{\Lambda_{\mathrm{C}}} \equiv 
\biggl(\frac{T}{T_{\mathrm{e}\gamma}}\biggr)^{-5/2},
\label{R1}\\
\frac{3 \rho_{\mathrm{e}} \, \Gamma_{\mathrm{e\, i}}}{4 \, \rho_{\gamma} \Gamma_{\gamma\mathrm{i}}} &=& 
\biggl(\frac{m_{\mathrm{p}}}{m_{\mathrm{e}}}\biggr)^2 \,\biggl(\frac{T}{T_{\mathrm{e}\gamma}}\biggr)^{-5/2} \equiv 
\biggl(\frac{T}{T_{\mathrm{i}\gamma}}\biggr)^{-5/2},
\label{R2}
\end{eqnarray}
where $\zeta(3) =1.202...$ and the ion mass has been estimated through the proton mass; the effective temperatures 
$T_{\mathrm{e}\gamma}$ and $T_{\mathrm{i}\gamma}$ introduced in the second equality of Eqs. (\ref{R1}) 
and (\ref{R2}) are defined as:
\begin{equation}
T_{\mathrm{e}\gamma} = m_{\mathrm{e}} \, {\mathcal N}^{2/5} \, \eta_{\mathrm{b}0}^{2/5}, \qquad 
T_{\mathrm{i}\gamma} = m_{\mathrm{e}}^{-1/5} m_{\mathrm{p}}^{4/5}  {\mathcal N}^{2/5} \, \eta_{\mathrm{b}0}^{2/5},
\qquad {\mathcal N} = \frac{270 \zeta(3)}{32 \, \pi^5} \ln{\Lambda_{\mathrm{C}}}.
\label{R3}
\end{equation}
In explicit terms and for the fiducial set of cosmological parameters determined on the basis of the WMAP 7yr data 
alone in the light of the $\Lambda$CDM scenario \cite{wmap7a,wmap7b}
\begin{equation}
T_{\mathrm{e}\gamma} = 88.6\, \biggl(\frac{h_{0}^2 \Omega_{\mathrm{b}0}}{0.02258}\biggr)^{2/5} \, \mathrm{eV},\qquad 
T_{\mathrm{i}\gamma} = 36.08\, \biggl(\frac{h_{0}^2 \Omega_{\mathrm{b}0}}{0.02258}\biggr)^{2/5} \, \mathrm{keV}.
\label{R4}
\end{equation}
On the basis of Eq. (\ref{R4}) there are three different dynamical regimes. When $T> T_{\mathrm{i}\gamma}$ the 
ion-photon and the electron-photon rates dominate against the Coulomb rate: in this regime the photons, electrons 
and ions are all coupled together and form a unique physical fluid with the same effective velocity. When 
$T_{\mathrm{e}\gamma}< T < T_{\mathrm{i}\gamma}$ the electron-photon rate dominates against the Coulomb 
rate which is anyway larger than the ion-photon rate. Finally for $T< T_{\mathrm{e} \gamma}$ the Coulomb 
rate is always dominant which means that the ion-electron fluid represents a unique entity characterized 
by a single velocity which is customarily referred to as the baryon velocity. 
The effective temperatures $T_{\mathrm{e}\gamma}$ and $T_{\mathrm{e\, i}}$ determine the hierarchies 
between the different rates and should not be confused with the kinetic temperatures of the electrons and of the ions 
which coincide approximately with the photon temperature $T_{\gamma} \simeq T_{\mathrm{e}} \simeq T_{\mathrm{i}}$.
For instance after matter-radiation equality $(T_{\mathrm{e}} - T_{\gamma})/T_{\gamma} \simeq {\mathcal O}(H/\Gamma_{\mathrm{e}\gamma})$ and $(T_{\mathrm{i}} - T_{\mathrm{e}})/T_{\gamma} \simeq 
{\mathcal O}(H/\Gamma_{\mathrm{e}\mathrm{i}})$  where $H$ is the standard Hubble rate at the corresponding epoch. 

Depending on the range of temperatures the effective evolution equations for the vorticities will change. 
In the regime $T> T_{\mathrm{i}\gamma}$ the Coulomb rate can be neglected in comparison 
with the Thomson rates and the vorticities of photons, electrons and ions approximately 
coincide. For $T_{\mathrm{e}\gamma}< T < T_{\mathrm{i}\gamma}$ the Ohm law can be easily 
obtained from Eq. (\ref{SA}) and it is given by 
\begin{equation}
\vec{E} + \vec{v}_{\mathrm{e}} \times \vec{B} = \frac{\vec{J}}{\sigma} + 
\frac{4}{3} \frac{\rho_{\gamma}}{\rho_{\mathrm{b}}} \frac{m_{\mathrm{i}}}{e} a^2 \Gamma_{\gamma\mathrm{e}}(\vec{v}_{\gamma}
- \vec{v}_{\mathrm{e}}),
\label{dc5}
\end{equation}
where it has been used that the baryon density $\rho_{\mathrm{b}} = (m_{\mathrm{i}} + m_{\mathrm{e}}) \tilde{n}_{0}$
coincides approximately with the ion density in the globally neutral case and that $n_{0} = a^3 \tilde{n}_{0}$; furthermore, in Eq. (\ref{dc5}), $\sigma$ denotes 
the electric conductivity \cite{mg1aa}
\begin{equation}
\sigma = \frac{\omega_{\mathrm{p\, e}}^2}{4 \pi a \Gamma_{\mathrm{e i}}}, \qquad \omega_{\mathrm{p}\, e} = \sqrt{\frac{4 \pi e^2\, n_{\mathrm{e}}}{m_{\mathrm{e}} a}},
\label{dc5aa}
\end{equation}
expressed in terms of the Coulomb rate and in terms of the electron 
plasma frequency\footnote{The electron plasma frequency of Eq. (\ref{dc5aa})
must not be confused with the vorticity} $\omega_{\mathrm{p\,e}}$. By taking the curl of both sides of Eq. (\ref{dc5}) the following relation can be easily derived:
\begin{equation}
\vec{\nabla}\times \vec{E} + \vec{\nabla} \times (\vec{v}_{\mathrm{e}}\times \vec{B})= 
\frac{\vec{\nabla}\times \vec{J}}{\sigma} + \frac{4}{3} \frac{\rho_{\gamma}}{\rho_{\mathrm{b}}} \frac{m_{\mathrm{i}}}{e}
a^2 \Gamma_{\gamma\mathrm{e}} (\vec{\omega}_{\gamma} - \vec{\omega}_{\mathrm{e}}).
\label{dc6}
\end{equation}
Recalling now Eq. (\ref{S2}) and (\ref{S3}), Eq. (\ref{dc6}) becomes:  
\begin{eqnarray}
\frac{\partial \vec{B}}{\partial \tau} = \vec{\nabla} \times (\vec{v}_{\mathrm{e}} \times \vec{B}) + \frac{\nabla^2 \vec{B}}{4\pi \sigma} - \frac{4}{3} \frac{\rho_{\gamma}}{\rho_{\mathrm{b}}} \, a^2 \, \frac{m_{\mathrm{i}}}{e} \Gamma_{\mathrm{e}\gamma} (\vec{\omega}_{\gamma} - \vec{\omega}_{\mathrm{e}}).
\label{dc6a}
\end{eqnarray}
In the same regime the evolution equation for the vorticities of the ions and of the photons are, up to spatial gradients, 
\begin{eqnarray}
&& \partial_{\tau} \vec{\omega}_{\mathrm{i}} + {\mathcal H} \vec{\omega}_{\mathrm{i}} = 
- \frac{e n_{\mathrm{i}}}{\rho_{\mathrm{i}} a^4} \partial_{\tau} \vec{B},
\label{dc7}\\
&& \partial_{\tau} \vec{\omega}_{\gamma} =a \Gamma_{\gamma\mathrm{e}} ( \vec{\omega}_{\mathrm{e}}-\vec{\omega}_{\gamma}).
\label{dc8}
\end{eqnarray}
By eliminating the electron-photon rate between Eqs. (\ref{dc7}) and (\ref{dc8}) and by neglecting the 
spatial gradients in Eq. (\ref{dc6a}), the following pair of approximate conservation laws can be obtained 
\begin{eqnarray}
\partial_{\tau} \biggl( a \vec{\omega}_{\mathrm{i}} + \frac{e}{m_{\mathrm{i}}} \vec{B}\biggr) =0,
\label{dc10a}\\
\partial_{\tau} \biggl( \frac{e}{m_{\mathrm{i}}} \vec{B} - \frac{a}{R_{\mathrm{b}}} \vec{\omega}_{\gamma}\biggr) =0,
\label{dc10b}
\end{eqnarray}
where the ratio $R_{\mathrm{b}}$ is given by:
\begin{equation}
R_{\mathrm{b}} = \frac{3}{4} \frac{\rho_{\mathrm{b}}}{\rho_{\gamma}} = 30.36 \, \biggl(\frac{10^{3}}{z}\biggr) \, h_{0}^2
\Omega_{\mathrm{b}0}.
\label{dc11}
\end{equation}
By further combining the relations of Eqs. (\ref{dc10a}) and (\ref{dc10b}) the vorticity of the photons can be directly related 
to the vorticity of the ions since $\partial_{\tau}[ R_{\mathrm{b}} \vec{\omega}_{\mathrm{i}} + \vec{\omega}_{\gamma}] =0$. By assuming that 
at a given time $\tau_{\mathrm{r}}$ the primordial value of the vorticity in the 
electron photon system  is $\vec{\omega}_{\mathrm{r}}$ and that $\vec{B}(\tau_{\mathrm{r}})=0$
we shall have that 
\begin{equation}
a_{\mathrm{r}} \vec{\omega}_{\mathrm{i}}(\tau_{\mathrm{r}}) + 
\frac{4}{3} \frac{\rho_{\gamma}(\tau_{\mathrm{r}})}{\rho_{\mathrm{b}}(\tau_{\mathrm{r}})} 
a_{\mathrm{r}}  \vec{\omega}_{\gamma}(\tau_{\mathrm{r}}) = \vec{\omega}_{\mathrm{r}}.
\label{dc11a}
\end{equation}
Thus the solution of Eqs. (\ref{dc10a}) and (\ref{dc10b}) with the initial condition (\ref{dc11a}) can be written as:
\begin{eqnarray}
&& \vec{\omega}_{\mathrm{i}}(\vec{x},\tau) = - \frac{e}{m_{\mathrm{i}}} \frac{\vec{B}(\vec{x},\tau)}{a(\tau)} + \frac{a_{\mathrm{r}}}{a(\tau)} \vec{\omega}_{\mathrm{r}},
\label{dc12}\\
&& \vec{\omega}_{\gamma}(\vec{x},\tau) = \frac{R_{\mathrm{b}}(\tau)}{a(\tau)} [ \vec{\omega}_{\mathrm{r}} - a(\tau) 
\vec{\omega}_{\mathrm{i}}(\vec{x},\tau)].
\label{dc13}
\end{eqnarray}
The approximate conservation laws of Eqs. (\ref{dc10a})--(\ref{dc10b}) can also be phrased in terms of the physical vorticities 
$\vec{\Omega}_{X}(\vec{x},\tau) = a(\tau) \vec{\omega}_{X}(\vec{x},\tau)$ where $X$ denotes a generic subscript\footnote{Note that while $\vec{\omega}_{X}$ is related to $\vec{B}$, the physical vorticity $\vec{\Omega}_{X}$ is directly proportional to $\vec{{\mathcal B}}$. For instance, in the treatment of  \cite{harrison1,harrison2,harrison3} the use of the physical vorticity and of the physical magnetic field is preferred.}. 

For typical temperatures $T< T_{\mathrm{e}\gamma}$ the electrons and the ions are more strongly coupled than the 
electrons and the photons. This means that the effective evolution can be described in terms of the one-fluid magnetohydrodynamical 
(MHD in what folllows) equations  where, on top of the total current $\vec{J}$  the 
center of mass vorticity of the electron-ion system is introduced 
\begin{equation}
\vec{\omega}_{\mathrm{b}} = \frac{m_{\mathrm{i}} \vec{\omega}_{\mathrm{i}} + m_{\mathrm{e}} \vec{\omega}_{\mathrm{e}}}{m_{\mathrm{e}} + m_{\mathrm{i}}}.
\label{dc3}
\end{equation}
Equation (\ref{S6}) (multiplied by $m_{\mathrm{e}}$) and Eq. (\ref{S7}) (multiplied 
by $m_{\mathrm{i}}$) can therefore be summed up with the result that 
\begin{equation}
\partial_{\tau}\vec{\omega}_{\mathrm{b}} + {\mathcal H}  \vec{\omega}_{\mathrm{b}} = \frac{\vec{\nabla}\times(\vec{J}\times 
\vec{B})}{a^4 \rho_{\mathrm{b}}} + \frac{4}{3}
 \frac{\rho_{\gamma}}{\rho_{\mathrm{b}}} a \Gamma_{\gamma\, \mathrm{e}} (\vec{\omega}_{\gamma} - \vec{\omega}_{\mathrm{b}}).
\label{dc4}
\end{equation}
The evolution equation for the total current can be obtained from the difference of Eqs. (\ref{SA}) and (\ref{SB}). Since 
the interaction rates are typically much larger than the expansion rates the Ohm equation can be simplified and becomes 
\begin{equation}
\vec{E} + \vec{v}_{\mathrm{b}} \times \vec{B} = \frac{\vec{J}}{\sigma} + 
\frac{4}{3} \frac{\rho_{\gamma}}{\rho_{\mathrm{b}}} \frac{m_{\mathrm{i}}}{e} a^2 \Gamma_{\gamma\mathrm{e}}(\vec{v}_{\gamma}
- \vec{v}_{\mathrm{b}}),
\label{dc4a}
\end{equation}
where $\vec{v}_{\mathrm{b}}$ is the baryon velocity related to the baryon vorticity as 
$\vec{\omega}_{\mathrm{b}} = \vec{\nabla} \times \vec{v}_{\mathrm{b}}$.
The similarity of Eqs. (\ref{dc6}) and (\ref{dc4})  should not be 
misunderstood: while Eq. (\ref{dc6}) follows from the right hand side 
of Eq. (\ref{SA}), Eq. (\ref{dc4}) follows by taking the difference of Eq. (\ref{SB}) (multiplied by $n_{\mathrm{i}}$) and of Eq. (\ref{SA}) (multiplied by $n_{\mathrm{e}}$).
The expression obtained by means of the latter difference is rather 
lengthy and can be found in its full generality, in Ref.  \cite{mg1aa} (see, in particular, Eqs. (7) and (10)). Here the expression has been simplified by neglecting 
higher orders in $(m_{\mathrm{e}}/m_{\mathrm{i}})$.
The effective set of evolution equations can then be written, in this regime, as 
\begin{eqnarray}
&& \partial_{\tau} \vec{\omega}_{\mathrm{b}} + {\mathcal H} \vec{\omega}_{\mathrm{b}} = 
\frac{\vec{\nabla}\times(\vec{J} \times \vec{B})}{a^4 \, \rho_{\mathrm{b}}} + \frac{\epsilon'}{R_{\mathrm{b}}} 
(\vec{\omega}_{\gamma} - \vec{\omega}_{\mathrm{b}}),
\label{dc4b}\\
&& \partial_{\tau} \vec{B} = \vec{\nabla}\times(\vec{v}_{\mathrm{b}}\times \vec{B}) + \frac{\nabla^2 \vec{B}}{4 \pi \sigma} 
+ \frac{m_{\mathrm{i}} a}{e \, R_{\mathrm{b}}} \epsilon' (\vec{\omega}_{\mathrm{b}} - \vec{\omega}_{\gamma}),
\label{dc4c}\\
&& \partial_{\tau} \vec{\omega}_{\gamma} = \epsilon' (\vec{\omega}_{\mathrm{b}} - \vec{\omega}_{\gamma}),
\label{dc4d}
\end{eqnarray}
where $\epsilon' = a \Gamma_{\mathrm{e}\gamma}$ is the differential optical depth where, as usual, the 
contribution of the ions has been neglected. In the tight coupling limit Eqs. (\ref{dc4b}), (\ref{dc4c}) and (\ref{dc4d}) 
imply that $\vec{\omega}_{\mathrm{b}\gamma} \simeq \vec{\omega}_{\mathrm{b}} \simeq \vec{\omega}_{\gamma}$ 
while $\vec{\omega}_{\mathrm{b}\gamma}$ obeys 
\begin{equation}
\partial_{\tau} \vec{\omega}_{\mathrm{b}\gamma} + \frac{{\mathcal H} R_{\mathrm{b}}}{R_{\mathrm{b}} + 1} 
\vec{\omega}_{\mathrm{b}\gamma} = R_{\mathrm{b}}\frac{\vec{\nabla}\times (\vec{J} \times \vec{B})}{\rho_{\mathrm{b}} \, a^4
 (R_{\mathrm{b}} + 1)}.
 \label{dcde}
 \end{equation}
In analogy with what has been done before, the conservation laws can be derived by combining Eqs. (\ref{dc4b}) and (\ref{dc4c}) 
\begin{equation}
\partial_{\tau} \biggl( \vec{B} + \frac{m_{\mathrm{i}}}{e}\, a \,\vec{\omega}_{\mathrm{b}}\biggr)  = 
\vec{\nabla} \times (\vec{v}_{\mathrm{b}} \times \vec{B}) + \frac{\nabla^2 \vec{B}}{4\pi \sigma} + 
\frac{m_{\mathrm{i}}}{e} \frac{\vec{\nabla} \times (\vec{J} \times \vec{B})}{a^3 \rho_{\mathrm{b}}}.
\label{dcdf}
\end{equation}
From Eqs. (\ref{dc4c}) and (\ref{dc4d}) and by neglecting the spatial 
gradients it also follows 
\begin{equation}
\partial_{\tau} \biggl( \vec{B} - \frac{a}{R_{\mathrm{b}}} \frac{m_{\mathrm{i}}}{e} \vec{\omega}_{\gamma} \biggr) = 0.
\label{dcdg}
\end{equation}
Equations (\ref{dcdf}) and (\ref{dcdg})
are separately valid, but, taken together and in the limit of tight baryon-photon coupling, 
they imply that the magnetic filed must be zero when the tight-coupling is exact (i.e. $\vec{\omega}_{\gamma} =\vec{\omega}_{\mathrm{b}}$).
In spite of the various physical regimes encountered in the analysis of the evolution of the vorticity the key point is to find a suitable source of large-scale vorticity which could be converted, in some way into a large-scale magnetic field \cite{reviewmax} (see also \cite{cov1,cov2}).  The conversion can not only occur prior to matter-radiation equality but also after \cite{mishustin} in the 
regime where, as explained,  the baryon-photon coupling becomes weak. Indeed, Eqs. (\ref{dc10a}) and (\ref{dcdf}) have the same 
dynamical content when the spatial gradients are neglected and the only difference involves the coupling to the photons.

There have been, through the years, suggestions involving primordial turbulence 
(see the interesting accounts of Refs. \cite{B1a}),  cosmic strings  with small scale structure (see, e. g. \cite{vort1,shellard1,dm2}). Since 
matter flow in baryonic wakes is turbulent, velocity gradients will be induced in the flow by the small-scale wiggles of the string producing ultimately the vorticity. Dynamical friction between cosmic strings and matter may provide a further source of vorticity \cite{shellard1}. There have been also studies 
trying to generate large-scale magnetic fields in the context of superconducting cosmic strings (see, for instance,\cite{dm2} and references therein). The possible generation of large-scale magnetic fields prior to hydrogen 
recombination has been discussed in \cite{dolr1,dolr2,hoganr} (see also \cite{dolr3}). The vorticity required in order to produce the magnetic fields is generated, according to \cite{dolr1}, by the photon  diffusion at second order in the temperature fluctuations. In a similar perspective Hogan \cite{hoganr} got less optimistic estimates which, according to \cite{dolr1,dolr2}, should be attributed to different approximation schemes employed in the analysis. Along this perspective various analyses 
discussed higher-order effects using the conventional 
perturbative expansion in the presence of the relativistic fluctuations 
of the geometry \cite{vort2}.  In the present paper, as already mentioned, we are going to follow a different route 
since we intend to use the gradient expansion for a direct estimate of the vorticity.

\renewcommand{\theequation}{3.\arabic{equation}}
\setcounter{equation}{0}
\section{Vorticity evolution in gradient expansion}
\label{sec3}
The conservation laws derived in section \ref{sec2} hold under the hypothesis that the spatial gradients are neglected in the evolution equations of the vorticity.
The logic of the gradient expansion \cite{gr1,gr2,gr3,gr4,gr5,gr6} can be combined 
with the tenets of the drift approximation  \cite{gr7,gr8,gr9} in the context of the ADM decomposition \cite{ADM1,ADM2}. It will be shown hereunder that the resulting formalism \cite{mg2} provides a more general description of the angular 
momentum transfer between the various species of the plasma. 
Consider therefore the standard ADM decomposition where the shift vectors are set to zero but the lapse function kept arbitrary, i.e. $g_{00}(\vec{x},\tau) = N^2(\vec{x},\tau)$ and $g_{ij}(\vec{x},\tau) = - \gamma_{ij}(\vec{x},\tau)$. In this case the Maxwell equations can be written  as
\begin{eqnarray}
&& \vec{\partial} \cdot \vec{E} = 4 \pi e [n_{\mathrm{i}} - n_{\mathrm{e}}], 
\qquad \vec{\partial} \cdot \vec{B}=0,
\label{NH1}\\
&&\partial_{\tau} \vec{B} + \vec{\partial} \times \vec{E}  =0,\qquad  \vec{\partial} \times \vec{B} = 4 \pi e \biggl[n_{\mathrm{i}} \, \vec{v}_{\mathrm{i}} - n_{\mathrm{e}} \, \vec{v}_{\mathrm{e}}\biggr] + \partial_{\tau}\vec{E},
\label{NH2}
\end{eqnarray}
where the rescaled electric and magnetic fields are given by:
\begin{equation}
E^{i}(\vec{x},\tau) = \biggl(\frac{\sqrt{\gamma}}{N}\biggr)_{(\vec{x},\tau)} {\mathcal E}^{i}(\vec{x},\tau),\qquad B^{i}(\vec{x},\tau) = \biggl(\frac{\sqrt{\gamma}}{N}\biggr)_{(\vec{x},\tau)} {\mathcal B}^{i}(\vec{x},\tau);
\label{NH3}
\end{equation}
in Eq. (\ref{NH3})  the subscripts specify that the rescaling is space-time dependent. The rescaled concentrations are 
\begin{equation}
n_{\mathrm{i}}(\vec{x},\tau) = \sqrt{\gamma} \, \tilde{n}_{\mathrm{i}}(\vec{x},\tau), 
\qquad n_{\mathrm{e}}(\vec{x},\tau) = \sqrt{\gamma} \, \tilde{n}_{\mathrm{e}}(\vec{x},\tau).
\label{NH3a}
\end{equation}
The shorthand notation\footnote{Note that the operators introduced in Eqs. (\ref{NH1})--(\ref{NH3}) are the generalized curl, divergence and gradient operators; they reduce to the conventional curl, divergence and gradient operators in the conformally flat limit.} employed in Eqs. (\ref{NH1})--(\ref{NH3}) implies
for a generic vector $A^{i}$, 
\begin{equation} 
\vec{\partial}\cdot \vec{A} \equiv \partial_{i} A^{i},\qquad 
(\vec{\partial}\times \vec{A})^{i} = \partial_{j}\biggl[N \gamma^{i k} \, \gamma^{j n} \, \eta_{n m k} \, A^{m}\biggr].
\label{NH4}
\end{equation}
In appendix \ref{APPA} some relevant complements on this formalism have been collected to avoid 
a digression from the main line of arguments contained in the present section. Two relevant 
aspects must anyway be borne in mind:
\begin{itemize}
\item{} in the conformally flat limit (i.e. $N(\vec{x}, \tau) \to a(\tau)$ and $\gamma_{ij}(\vec{x}, \tau) \to 
a^2(\tau) \delta_{ij}$) Eqs. (\ref{NH1}) and (\ref{NH2}) reproduce exactly Eqs. (\ref{S1}) and (\ref{S2});
\item{} the same comment holds for all the other fields (i.e. comoving or physical) involved 
in the fully inhomogeneous description.
\end{itemize}
Using the generalized curl operator of Eq. (\ref{NH4}) the vorticity of the ions, of the electrons and of the 
photons can be written as 
\begin{equation}
\omega^{i}_{\mathrm{i}} = \partial_{j} \bigl( \Lambda^{ij}_{m} \, v_{\mathrm{i}}^{m}\bigr),\qquad 
\omega^{i}_{\mathrm{e}} = \partial_{j} \bigl( \Lambda^{ij}_{m} \, v_{\mathrm{e}}^{m}\bigr),
\qquad 
\omega^{i}_{\gamma} = \partial_{j} \bigl( \Lambda^{ij}_{m} \, v_{\gamma}^{m}\bigr),
\label{NH5}
\end{equation}
where $K_{ij}$ is the extrinsic curvature (see appendix \ref{APPA}) while $\Lambda_{m}^{ij}$ and $\overline{\Lambda}^{ij}_{m}$ are defined as\footnote{Recall that $\eta_{a b c} = \sqrt{\gamma} \, \epsilon_{a b c}$ and that $\eta^{a b c} = \epsilon^{a b c}/\sqrt{\gamma}$.}: 
\begin{equation}
\Lambda^{ij}_{m} = N \gamma^{ik}\,\gamma^{j n} \, \eta_{n m k},\qquad 
\overline{\Lambda}^{ij}_{m} = 2 N^2 [ K^{i k} \, \gamma^{j n} + K^{j n} \, \gamma^{i k} ] \eta_{n m k}.
\label{NH5a}
\end{equation}
Using Eqs. (\ref{NH5})--(\ref{NH5a}) as well as the evolution equations of the velocities (see, Eqs. (\ref{Av1})--(\ref{Av2})), the  evolution for the vorticity of the electrons and of the ions can be written, respectively, as\footnote{We shall focus, without loss of generality, on the situation where the lapse function is homogeneous, i.e. $N(\vec{x},\tau) = N(\tau)$; in this case 
the already lengthy expressions will be more manageable since the spatial derivatives of the lapse function 
will vanish.} 
\begin{eqnarray}
&& \partial_{\tau} \omega_{\mathrm{e}}^{i} + \biggl( N K - \frac{\partial_{\tau} N}{N}\biggr) \omega_{\mathrm{e}}^{i} - 
{\mathcal G}^{i}_{k} \omega^{k}_{\mathrm{e}} - {\mathcal F}_{\mathrm{e}}^{i} =
\nonumber\\
&&-  \frac{e \tilde{n}_{\mathrm{e}} N^2}{ \rho_{\mathrm{e}} \sqrt{\gamma}} \biggl\{ (\vec{\partial} \times \vec{E})^{i} + [ \vec{\partial}\times (\vec{v}_{\mathrm{e}} \times \vec{B})]^{i} \biggr\} + N \Gamma_{\mathrm{e}\mathrm{i}} ( \omega^{i}_{\mathrm{i}} - 
\omega^{i}_{\mathrm{e}}) 
+ \frac{4}{3} \frac{\rho_{\gamma}}{\rho_{\mathrm{e}}} N \Gamma_{\mathrm{e} \gamma}(\omega_{\gamma}^{i} - \omega_{\mathrm{e}}^{i}),
\label{NH6}\\
&& \partial_{\tau} \omega_{\mathrm{i}}^{i} + \biggl( N K - \frac{\partial_{\tau} N}{N}\biggr) \omega_{\mathrm{i}}^{i} - 
{\mathcal G}^{i}_{k} \omega^{k}_{\mathrm{i}} - {\mathcal F}_{\mathrm{i}}^{i} =
\nonumber\\
&&  \frac{e \tilde{n}_{\mathrm{i}} N^2}{ \rho_{\mathrm{i}} \sqrt{\gamma}} \biggl\{ (\vec{\partial} \times \vec{E})^{i} + [ \vec{\partial}\times (\vec{v}_{\mathrm{i}} \times \vec{B})]^{i} \biggr\} + N \Gamma_{\mathrm{i}\mathrm{e}} \frac{\rho_{\mathrm{e}}}{\rho_{\mathrm{i}}}( \omega^{i}_{\mathrm{e}} - 
\omega^{i}_{\mathrm{i}}) 
+ \frac{4}{3} \frac{\rho_{\gamma}}{\rho_{\mathrm{i}}} N \Gamma_{\mathrm{i} \gamma}(\omega_{\gamma}^{i} - \omega_{\mathrm{i}}^{i}).
\label{NH7}
\end{eqnarray}
Similarly, from Eq. (\ref{Av2a}) the evolution equation for the vorticity of the photons can be written as
\begin{equation}
\partial_{\tau} \omega_{\gamma}^{i} + \biggl[ \frac{4}{3} N K - \frac{\partial_{\tau} N}{N} \biggr] \omega_{\gamma}^{i} - 
{\mathcal G}^{i}_{k} \omega_{\gamma}^{k} - {\mathcal F}^{i}_{\gamma} 
 = N \Gamma_{\gamma\mathrm{e}} (\omega_{\mathrm{e}}^{i} - 
\omega_{\gamma}^{i}) + N \Gamma_{\gamma\mathrm{i}} ( \omega_{\mathrm{i}}^{i} - \omega_{\gamma}^{i}).
\label{NH7a}
\end{equation}
The quantities ${\mathcal F}_{\mathrm{e}}^{i}$, 
${\mathcal F}_{\mathrm{i}}^{i}$ and ${\mathcal F}_{\gamma}^{i}$ appearing in Eqs. (\ref{NH6}), (\ref{NH7}) and (\ref{NH7a}) are of the same order of the other terms appearing in the equations and they are defined as 
\begin{eqnarray}
{\mathcal F}_{\mathrm{e}}^{i} &=& \partial_{j} \biggl( \overline{\Lambda}^{ij}_{m} v^{m}_{\mathrm{e}}\biggr) + 
\frac{4}{3} N \Gamma_{\gamma\mathrm{e}} \partial_{j} \biggl(\frac{\rho_{\gamma}}{\rho_{\mathrm{e}}}\biggr) \Lambda^{ij}_{m}
(v^{m}_{\gamma} - v^{m}_{\mathrm{e}}), 
\nonumber\\
&+& \partial_{j} {\mathcal G}^{m}_{a} \Lambda^{ij}_{m} v_{\mathrm{e}}^{a} - N \partial_{j} K \Lambda^{ij}_{m} v^{m}_{\mathrm{e}} - \partial_{j} \biggl( \frac{e \tilde{n}_{\mathrm{e}} N^2}{\rho_{\mathrm{e}} \sqrt{\gamma}} \biggr) \Lambda^{ij}_{m}\biggl[ E^{m} + (\vec{v}_{\mathrm{e}}\times \vec{B})^{m} \biggr],
\label{NH8}\\
{\mathcal F}^{i}_{\mathrm{i}} &=& \partial_{j} \biggl( \overline{\Lambda}^{ij}_{m} v^{m}_{\mathrm{i}}\biggr) + 
\frac{4}{3} N \Gamma_{\gamma\mathrm{i}} \partial_{j} \biggl(\frac{\rho_{\gamma}}{\rho_{\mathrm{i}}}\biggr) \Lambda^{ij}_{m}
(v^{m}_{\gamma} - v^{m}_{\mathrm{i}}) + N \partial_{j}\biggl(\frac{\rho_{\mathrm{e}}}{\rho_{\mathrm{i}}}\biggr) \, \Lambda^{ij}_{m}\, \Gamma_{\mathrm{i e}} (v^{m}_{\mathrm{e}} - v^{m}_{\mathrm{i}}), 
\nonumber\\
&+& \partial_{j} {\mathcal G}^{m}_{a} \Lambda^{ij}_{m} v_{\mathrm{i}}^{a} - N \partial_{j} K \Lambda^{ij}_{m} v^{m}_{\mathrm{i}} + \partial_{j} \biggl( \frac{e \tilde{n}_{\mathrm{i}} N^2}{\rho_{\mathrm{i}} \sqrt{\gamma}} \biggr) \Lambda^{ij}_{m} \biggl[ E^{m} + (\vec{v}_{\mathrm{i}}\times \vec{B})^{m} \biggr],
\label{NH9}\\
{\mathcal F}_{\gamma}^{i} &=& \partial_{j} \biggl(\overline{\Lambda}^{i j}_{k} \, v^{k}_{\gamma}\biggr) + \Lambda^{i j}_{k} \, v^{q}_{\gamma} \partial_{j} {\mathcal G}^{k}_{q} - \frac{4}{3} N\partial_{j} K \, \Lambda^{i j}_{k} \, v^{k}_{\gamma}  - \frac{N^2}{4} \partial_{j}\biggl\{ \frac{\Lambda^{i j}_{k}}{\rho_{\gamma}} \partial_{m}\biggl[ \rho_{\gamma} \gamma^{m k} \biggr]\biggr\}.
\label{NH9a}
\end{eqnarray}
The generalized scalar and vector products appearing in Eqs. (\ref{NH8}), (\ref{NH9}) and (\ref{NH9a}) are defined as 
\begin{equation}
\vec{F} \cdot \vec{G} = \gamma_{m n} F^{m} G^{n},\qquad (\vec{F} \times \vec{G})^{k} = \frac{\gamma_{i n} \gamma_{m \ell}}{N}
F^{n} G^{m} \eta^{i \,\ell\, k},
\label{NH9b}
\end{equation}
and coincide with the ordinary scalar and vector products in the conformally 
flat limit introduced after Eq. (\ref{NH4}). 
The velocity fields appearing in Eqs. (\ref{NH6}) and (\ref{NH7}) are all subjected to the fully inhomogeneous 
form of the momentum constraint implying, from Eq. (\ref{0i}),
\begin{equation}
\frac{1}{N} \biggl( \nabla_{i} K - \nabla_{k} K^{k}_{i} \biggr) = \ell_{\mathrm{P}}^2 (p + \rho) u^{0} u_{i},\qquad u^{0} = \frac{1}{N}
\sqrt{1 + u^2},
\label{CON1}
\end{equation}
where $u^2 = u^{i} u^{j} \gamma_{ij}$ and where $u^{0}$ and $u^{i}$ can also be defined in terms of the total velocity field 
 $v^{i}$ which turns out to be the weighted sum of the velocity fields 
of the electrically charged and of the electrically neutral species, i.e. 
\begin{equation}
(p + \rho) v^{k} = \sum_{a} ( p_{\mathrm{a}} + \rho_{\mathrm{a}}) v^{k}_{\mathrm{a}} = 
 \rho_{\mathrm{e}} v^{k}_{\mathrm{e}} + \rho_{\mathrm{i}} v^{k}_{\mathrm{i}} + \frac{4}{3} \rho_{\gamma} v^{k}_{\gamma}
 + \frac{4}{3} \rho_{\nu} v^{k}_{\nu} + \rho_{\mathrm{c}} v_{\mathrm{c}}^{k},
\label{CON4}
\end{equation}
where the contribution of the cold dark matter particles and of the massless 
neutrinos has been also added.
The explicit connection between $u^{0}$, $u^{i}$ and $v^{i}$ is given by: 
\begin{equation}
u^{0} = \frac{\cosh{y}}{N}, \qquad u^{i} = \frac{v^{i}}{N} \cosh{y}, \qquad \cosh{y} = \frac{1}{\sqrt{1 - v^2/N^2}},
\label{CON2}
\end{equation}
where $v^2 = v^{i} v^{j} \gamma_{ij}$. In terms of $v^{i}$ and $v^2$ the momentum constraint of Eq. (\ref{CON1}) can also be written as 
\begin{equation}
\ell_{\mathrm{P}}^2 ( p + \rho) \frac{v^{i}}{N} =  \biggl(1 - \frac{v^2}{N^2} \biggr) \, \nabla_{k} \biggl(K^{k i} - K \gamma^{k i}\biggr).
\label{CON3}
\end{equation}

All the discussion of section \ref{sec2} can be generalized to the fully inhomogeneous 
case and we shall be particularly interested in the generalization 
of the conservation laws determining the angular momentum exchange 
between the various species. Consider then the situation where 
the electron-photon rate dominates against the Coulomb rate. In this 
case the fully inhomogeneous form of the Ohm law reads
\begin{equation}
- E^{k}  - (\vec{v}_{\mathrm{e}} \times \vec{B})^{k} + \frac{J^{k}}{\sigma} + 
\frac{4}{3\, e} \, \frac{\rho_{\gamma}}{\rho_{\mathrm{b}}} m_{\mathrm{i}} \Gamma_{\mathrm{e} \gamma} 
\, \frac{\sqrt{\gamma}}{N} ( v_{\gamma}^{k} - v_{\mathrm{e}}^{k}) =0.
\label{D1}
\end{equation}
By taking  the generalized curl of Eq. (\ref{D1}) (see Eq. (\ref{NH4})) the following 
equation can be obtained
\begin{eqnarray}
&& - \vec{\partial} \times \vec{E}  - \vec{\partial}\times(\vec{v}_{\mathrm{e}} \times \vec{B}) + \vec{\partial}\times(\vec{J}/\sigma) 
\nonumber\\
&& +
\frac{4}{3\, e} \, \frac{\rho_{\gamma}}{\rho_{\mathrm{b}}} m_{\mathrm{i}} \Gamma_{\mathrm{e} \gamma} 
\, \frac{\sqrt{\gamma}}{N} ( \vec{\omega}_{\gamma} - \vec{\omega}_{\mathrm{e}}) 
- \frac{4}{3} \,\frac{m_{\mathrm{i}}}{e} \,N^2 (\vec{v}_{\gamma} -\vec{v}_{\mathrm{e}}) \times \vec{\partial} 
\biggl[ \Gamma_{\mathrm{e}\gamma} \frac{\sqrt{\gamma}}{N} \, \frac{\rho_{\gamma}}{\rho_{\mathrm{b}}}\biggr]=0,
\label{D2}
\end{eqnarray}
where, consistently with Eq. (\ref{NH9b}), the last term at the left hand side is defined in terms of the generalized 
vector product and it vanishes exactly in the conformally flat limit. 
By assuming, as physically plausible prior to decoupling, that the conductivity is 
homogeneous, Eqs. (\ref{NH1}) and (\ref{NH2}) can be used inside 
Eq. (\ref{D2}) and the final equation will then be:
\begin{eqnarray}
&& \partial_{\tau} \vec{B} = \vec{\partial}  \times (\vec{v}_{\mathrm{e}}\times \vec{B}) 
- \frac{1}{4\pi \sigma} \vec{\partial} \times (\vec{\partial} \times \vec{B}) - 
\frac{4}{3\, e} \, \frac{\rho_{\gamma}}{\rho_{\mathrm{b}}} m_{\mathrm{i}} \Gamma_{\mathrm{e} \gamma} 
\, \frac{\sqrt{\gamma}}{N} ( \vec{\omega}_{\gamma} - \vec{\omega}_{\mathrm{e}}) 
\nonumber\\
&& + \frac{4}{3} N^2 \frac{m_{\mathrm{i}}}{e} (\vec{v}_{\gamma} -\vec{v}_{\mathrm{e}}) \times \vec{\partial} \biggl[ \Gamma_{\mathrm{e}\gamma} \frac{\sqrt{\gamma}}{N}\, \frac{\rho_{\gamma}}{\rho_{\mathrm{b}}}\biggr].
\label{D3}
\end{eqnarray}
Equation (\ref{D3}) reduces, in the conformally flat limit, to Eq. (\ref{dc6a}). 
The same logic can be applied in all the other 
derivations and the obtained result expanded to first order in the spatial gradients 
with the result that the generalized system for the evolution 
of the vorticities reads 
\begin{eqnarray}
&& \partial_{\tau} \omega_{\mathrm{i}}^{k} = \biggl( N K + 2 \frac{\partial_{\tau} N}{N} 
\biggr) \omega_{\mathrm{i}}^{k} - \frac{e \tilde{n}_{\mathrm{i}}}{\rho_{\mathrm{i}} \sqrt{\gamma}} \, N^2 \, \partial_{\tau} B^{k},
\label{D4}\\
&& \partial_{\tau} B^{k} = -\frac{4}{3\, e} \Gamma_{\mathrm{e}\gamma} \frac{\rho_{\gamma}}{\rho_{\mathrm{b}}} m_{\mathrm{i}} \frac{\sqrt{\gamma}}{N} (\omega_{\gamma}^{k} -
\omega_{\mathrm{e}}^{k}),
\label{D5}\\
&& \partial_{\tau} \omega_{\gamma}^{k} = \biggl( \frac{2}{3} N K + 2 \frac{\partial_{\tau} N}{N} \biggr) \omega_{\gamma}^{k} + N \Gamma_{\mathrm{e}\gamma} 
(\omega_{\mathrm{e}}^{k} - \omega_{\gamma}^{k}).
\label{D6}
\end{eqnarray}
Equations (\ref{D4}), (\ref{D5}) and (\ref{D6}) reduce, respectively,  to Eqs. 
(\ref{dc6a}), (\ref{dc7}) and (\ref{dc8}) in the conformally flat limit. 
Equations (\ref{D4}), (\ref{D5}) and (\ref{D6}) apply in the situation  where
the magnetic fields are initially zero and do not contribute to the extrinsic curvature so that 
 $K_{i}^{j} = K/3 \delta_{i}^{j} + \overline{K}_{i}^{j}$ with $\overline{K}_{i}^{j}=0$.  In this case Eqs. (\ref{D4})--(\ref{D6}) reduce to a pair 
of remarkable conservation laws whose explicit expression, up to 
spatial gradients, is
\begin{eqnarray}
&&\partial_{\tau} \biggl[ \frac{\sqrt{\gamma}}{N^2} \omega_{\mathrm{i}}^{k} + 
\frac{ e \tilde{n}_{\mathrm{i}}}{\rho_{\mathrm{b}}} B^{k} \biggr] =0,
\label{D7}\\
&& \partial_{\tau} \biggl[ B^{k} - \frac{m_{\mathrm{i}}}{e \, \overline{R}_{\mathrm{b}}} 
\frac{\gamma^{1/3}}{N^2} \omega_{\gamma}^{k} \biggr] =0,
\label{D8}
\end{eqnarray}
where $\overline{R}_{\mathrm{b}}(\vec{x},\tau_{1})$ is a constant in time (but not in space) and come from the inhomogeneous generalization of $R_{\mathrm{b}}(\vec{x},\tau)$: 
\begin{equation}
R_{\mathrm{b}}(\vec{x},\tau) = \frac{3}{4} \frac{\rho_{\mathrm{b}}(\vec{x},\tau)}
{\rho_{\gamma}(\vec{x},\tau)} = \overline{R}_{\mathrm{b}}(\vec{x},\tau_{1})\gamma^{1/6}.
\label{D9}
\end{equation}
The evolution of the vorticity of the baryons 
as well as the tight coupling between the baryons and the photons can be discussed 
in full analogy with the considerations already developed above in the 
case of the electron-photon coupling. The inhomogeneous generalization 
of the Ohm law when the Coulomb scattering dominates against both the electron-photon and the ion-photon coupling has been derived in Ref. \cite{mg2} 
(see Eq. (3.34)). To leading order in the gradient expansion 
the evolution of the baryon vorticity can be written as
\begin{eqnarray}
&& \partial_{\tau} \omega_{\mathrm{b}}^{k} = \biggl( N K + 2 \frac{\partial_{\tau} N}{N} \biggr) \omega_{\mathrm{b}}^{k} + \frac{\epsilon'}{R_{b}}(\omega_{\gamma}^{k} - \omega_{\mathrm{b}}^{k}), 
\label{D10}\\
&& \partial_{\tau} B^{k} = - \frac{m_{\mathrm{i}}}{e} \frac{\epsilon'}{R_{\mathrm{b}}} 
\frac{\sqrt{\gamma}}{N^2} (\omega_{\gamma}^{k} - \omega_{\mathrm{b}}^{k}).
\label{D11}
\end{eqnarray}
where $\epsilon'= N \Gamma_{\mathrm{e}\gamma}$ is the inhomogeneous 
generalization of the optical depth. By eliminating 
$\epsilon'$ between Eqs. (\ref{D10}) and (\ref{D11}) the following equation
\begin{equation}
\partial_{\tau} \biggl[ \frac{\sqrt{\gamma}}{N^2} \omega_{\mathrm{b}}^{k} + 
\frac{ e \tilde{n}_{\mathrm{i}}}{\rho_{\mathrm{b}}} B^{k} \biggr] =0
\label{D12}
\end{equation}
is readily obtained. Note that Eq. (\ref{D12}) coincides, up to spatial 
gradients and in the conformally flat limit, with Eq. (\ref{dcdf}). 
\renewcommand{\theequation}{4.\arabic{equation}}
\setcounter{equation}{0}
\section{Maximal vorticity induced by the geometry}
\label{sec4}
 In this paper the the expansion is organized not 
in terms of the relative magnitude of the gravitational and electromagnetic fluctuations but in terms 
of the number of gradients carried by each order of the expansion. 
From the momentum constraint (see  Eq. (\ref{CON3})), the total 
velocity field can be written, formally, 
\begin{eqnarray}
v^{i} &=& - \frac{N\, S^{i}}{ 2 S^2} \biggl[ 1 - \sqrt{1 + 4 S^2} \biggr] \simeq N S^{i} \biggl[ 1 - S^2 + {\mathcal O}(\epsilon^3)\biggr] + 
{\mathcal O}(\epsilon^4), 
\label{S0}\\
S^{i} &=& \frac{1}{\ell_{\mathrm{P}}^2 ( p + \rho)} 
\nabla_{k}\biggl( K^{k i} - K \gamma^{k i}\biggr),
\label{ES1}
\end{eqnarray}
where the orders of the expansion appearing in Eq. (\ref{S0}) are defined by the number 
of gradients. From Eqs. (\ref{S0}) and (\ref{NH5})--(\ref{NH5a}) the total vorticity can be written as
\begin{equation}
\omega_{\mathrm{tot}}^{i} = \partial_{j}\biggl\{ N \Lambda^{ij}_{m} S^{m} \biggl[ 1 - S^2 + {\mathcal O}(\epsilon^3)\biggr]\biggr\}.
\label{ES2}
\end{equation} 
To implement the gradient expansion let us parametrize the geometry as 
\begin{equation}
\gamma_{ij}(\vec{x},\tau) = a^2(\tau)[ \alpha_{ij}(\vec{x}) + \beta_{ij}(\vec{x},\tau) ],
\qquad \gamma^{ij}(\vec{x},\tau) = \frac{1}{a^2(\tau)}[ \alpha^{i j}(\vec{x}) - \beta^{ij}(\vec{x},\tau)].
\label{ES3}
\end{equation} 
and keep the lapse function homogeneous, i.e. $N(\tau) = a(\tau)$; $\alpha_{ij}(\vec{x})$ does not contain any spatial gradient while $\beta_{ij}(\vec{x},\tau)$ contains at least one spatial gradient.  The extrinsic curvature becomes:
\begin{equation}
K_{i}^{j} = - \biggl(\frac{{\mathcal H}}{a} \delta_{i}^{j} + \frac{1}{2} \frac{\partial_{\tau} \beta_{i}^{j}}{a} \biggr),\qquad 
K^{i k} = - \frac{1}{a^3}\biggl[{\mathcal H}( \alpha^{i k} - \beta^{ik}) + \frac{1}{2} \partial_{\tau} \beta^{ik} \biggr].
\label{ES4}
\end{equation}
Furthermore we have also that the spatial Christoffel are:
\begin{eqnarray}
\Gamma^{k}_{k a} &=& \partial_{a} \ln{\sqrt{\gamma}} = \frac{1}{2} \biggl( \frac{\partial_{a} \alpha}{\alpha} + \partial_{a} \beta\biggr),
\label{ES5}\\
\Gamma^{m}_{a b} &=& \frac{1}{2}\biggl[  \alpha^{m n} \lambda_{n a b}  + \alpha^{m n} \overline{\lambda}_{n a b} - 
\beta^{m n} \lambda_{n a b}\biggr],
\label{ES6}
\end{eqnarray}
where $\lambda_{n a b}$ and $\overline{\lambda}_{n a b}$
\begin{eqnarray}
\lambda_{n a b} &=& - \partial_{n} \alpha_{a b} + \partial_{b} \alpha_{n a} + \partial_{a}\alpha_{b n},
\label{ES7}\\
\overline{\lambda}_{n a b} &=& - \partial_{n} \beta_{a b} + \partial_{b} \beta_{n a} + \partial_{a}\beta_{b n}.
\label{ES8}
\end{eqnarray}
The relevant term appearing in the momentum constraint becomes then 
\begin{equation}
\nabla_{k}\biggl( K^{ k m} - K \gamma^{k m} \biggr) = \nabla_{k} K^{k m} + \frac{\alpha^{k m} \partial_{\tau} \partial_{k} \beta}{2 a^3},
\label{ES9}
\end{equation}
where 
\begin{eqnarray}
\nabla_{k} K^{k m} &=& - \frac{1}{a^3} \biggl[\frac{{\mathcal H}}{ 2 \alpha} (\partial_{a} \alpha) \alpha^{a m} 
+ {\mathcal H} \partial_{k} \alpha^{k m} + {\mathcal H} \frac{\alpha^{m i} \alpha^{a b}}{2} \lambda_{i a b}\biggr]
\nonumber\\
&+& \frac{1}{2a^3} \biggl[- \partial_{k} \partial_{\tau} \beta^{km}  - \frac{1}{2}\biggl(\partial_{\tau} \beta^{a b} \alpha^{m i}\biggr) \lambda_{i a b} 
- \biggl(\frac{\partial_{a} \alpha}{2 \alpha}\biggr) \partial_{\tau} \beta^{a m} 
\nonumber\\
&-& {\mathcal H} \partial_{a} \beta \alpha^{a m} 
+ {\mathcal H} \biggl(\frac{\partial_{a} \alpha}{\alpha}\biggr) \beta^{am} 
+ 2 {\mathcal H} \partial_{k} \beta^{k m}
- {\mathcal H} \alpha^{m i} \alpha^{a b} \overline{\lambda}_{i a b}
\nonumber\\
&+& {\mathcal H} \biggl( \alpha^{m i} \beta^{ab} + \alpha^{a b} \beta^{m i}\biggr) \lambda_{i a b} 
\biggr].
\label{ES10}
\end{eqnarray}
Equation (\ref{ES9}) can therefore be written as 
\begin{eqnarray} 
\nabla_{k}\biggl( K^{ k m} - K \gamma^{k m} \biggr) &=& - \frac{{\mathcal H}}{a^3} \biggl[\frac{1}{ 2 \alpha} (\partial_{a} \alpha) \alpha^{a m} 
+ \partial_{k} \alpha^{k m} + \frac{\alpha^{m i} \alpha^{a b}}{2} \lambda_{i a b}\biggr]
\nonumber\\
&+& \frac{1}{2a^3} \biggl\{\alpha^{k m} \partial_{k} \partial_{\tau} \beta - \partial_{k} \partial_{\tau} \beta^{km}  
\nonumber\\
&-& \biggl[\frac{1}{2}\biggl(\partial_{\tau} \beta^{a b} \alpha^{m i}\biggr) \lambda_{i a b} 
+ \biggl(\frac{\partial_{a} \alpha}{2 \alpha}\biggr) \partial_{\tau} \beta^{a m}\biggr]  
\nonumber\\
&+&  {\mathcal H}\biggl[2 \partial_{k} \beta^{k m} 
- \partial_{a} \beta \alpha^{a m} + \biggl(\frac{\partial_{a} \alpha}{\alpha}\biggr) \beta^{am} 
-  \alpha^{m i} \alpha^{a b} \overline{\lambda}_{i a b}
\nonumber\\
&+&  \biggl( \alpha^{m i} \beta^{ab} + \alpha^{a b} \beta^{m i}\biggr) \lambda_{i a b} \biggr]
\biggr\}.
\label{ES11}
\end{eqnarray}
The previous expression can also be recast in a more handy form:
\begin{equation}
\nabla_{k}\biggl( K^{ k m} - K \gamma^{k m} \biggr) = - \frac{{\mathcal H}}{a^3} {\mathcal Z}^{m}(\alpha) + 
\frac{1}{2 a^3} \biggl[ {\mathcal I}_{1}^{m}(\alpha,\beta) - {\mathcal I}_{2}^{m}(\alpha,\beta) + {\mathcal H} {\mathcal I}^{m}_{3}(\alpha,\beta)\biggr],
\label{ES12}
\end{equation}
where the three functionals of $\alpha_{ij}(\vec{x})$ and $\beta_{ij}(\vec{x},\tau)$ are defined as 
\begin{eqnarray}
{\mathcal Z}^{m}(\alpha) &=& \frac{1}{2} \frac{\partial_{a} \alpha}{\alpha} \alpha^{a m} + \partial_{q} \alpha^{q m} + 
\frac{\alpha^{m q} \alpha^{a b}}{2} \lambda_{q a b},
\label{ES13}\\
{\mathcal I}_{1}^{m}(\alpha,\beta) &=&\alpha^{q m} \partial_{q} \partial_{\tau} \beta - \partial_{\tau} \partial_{q} \beta^{q m},
\label{ES14}\\
{\mathcal I}_{2}^{m}(\alpha,\beta) &=& \frac{\alpha^{q m}}{2} (\partial_{\tau} \beta^{a b}) \lambda_{q a b} + \frac{\partial_{a} \alpha}{2 \alpha} \partial_{\tau} \beta^{a m},
\label{ES15}\\
{\mathcal I}_{3}^{m}(\alpha,\beta) &=& 2 \partial_{q} \beta^{q m} - (\partial_{a} \beta) \alpha^{a m} + \frac{\partial_{a} \alpha}{\alpha} 
\beta^{a m} + \lambda_{q a b} \biggl( \alpha^{q m} \beta^{a b} + \alpha^{a b} \beta^{m q} \biggr) 
\nonumber\\
&-& \alpha^{m q} \alpha^{a b} \overline{\lambda}_{q a b}. 
\label{ES16}
\end{eqnarray}
With the result of Eq. (\ref{ES12}) we can compute the first relevant part of the final expression, namely:
\begin{eqnarray}
&& N^2 \gamma^{a j} \gamma^{i n} \eta_{a m n} \nabla_{k}\biggl( K^{ k m} - K \gamma^{k m} \biggr) = 
\frac{\sqrt{\alpha}}{a^2} \biggl( 1 + \frac{\beta}{2}\biggr) \biggl\{ - {\mathcal H} \alpha^{k j} \alpha^{i n} {\mathcal Z}^{m}(\alpha) 
\epsilon_{k m n} 
\nonumber\\
&+& {\mathcal H} \biggl( \alpha^{k j} \beta^{i n} + \alpha^{i n} \beta^{k j}\biggr) {\mathcal Z}^{m}(\alpha) \epsilon_{k m n} 
+ \frac{\alpha^{k j} \alpha^{i n}}{2} \epsilon_{k m n} \biggl[{\mathcal I}_{1}^{m}(\alpha,\beta) - {\mathcal I}^{m}_{2}(\alpha,\beta) 
\nonumber\\
&+& {\mathcal H} {\mathcal I}_{3}^{m}(\alpha,\beta) \biggr]\biggr\}.
\label{ES17}
\end{eqnarray}
Recalling that, furthermore\footnote{We shall assume that 
 $w$, the dominant barotropic index of the fluid sources, is constant.} 
\begin{equation}
\ell_{\mathrm{P}}^2 ( p + \rho) a^2 = \frac{3 {\mathcal H}_{1}^2 ( 1 + w)}{\alpha^{(w +1)/2} \, ( 1 + \beta/2)^{w +1}} 
\biggl(\frac{a_{1}}{a}\biggr)^{3 w+ 1},\qquad \ell_{\mathrm{P}}^2 \overline{\rho}_{1} a^2_{1} = 3 {\mathcal H}_{1}^2.
\label{ES18}
\end{equation}
Putting all the various parts of the calculation together we have that, from Eq. (\ref{ES2}),
\begin{equation}
\omega_{\mathrm{tot}}^{i} =  \partial_{j} {\mathcal A}^{i j}, \qquad 
{\mathcal A}^{ij} = \frac{N^2 \, \gamma^{k j} \, \gamma^{i n}\, \eta_{k m n}}{\ell_{\mathrm{P}}^2 ( p + \rho)} \nabla_{a}\biggl(
K^{a m} - \gamma^{a m} K\biggr),
\label{ES19}
\end{equation}
then the quantity ${\mathcal A}^{ij}$ becomes:
\begin{eqnarray}
{\mathcal A}^{ij}(\alpha,\beta) &=& \frac{ \alpha^{(w + 2)/2}}{3 {\mathcal H}_{1}^2 (w + 1)} \biggl(\frac{a}{a_{1}}\biggr)^{3 w+1} 
\biggl\{ - {\mathcal H} \alpha^{k j} \alpha^{i n}  {\mathcal Z}^{m}(\alpha) \epsilon_{k m n}
\nonumber\\
&+& {\mathcal H} \biggl[ \alpha^{k j} \beta^{i n} + \alpha^{i n} \beta^{k j}\biggr] {\mathcal Z}^{m}(\alpha) \epsilon_{k m n}
+ \frac{\alpha^{k j} \alpha^{i n}}{2} \epsilon_{k m n} \biggl[ {\mathcal I}_{1}^{m}(\alpha,\beta) 
\nonumber\\
&-& {\mathcal I}^{m}_{2}(\alpha,\beta) + {\mathcal H} {\mathcal I}_{3}^{m}(\alpha,\beta) \biggr]
- \frac{{\mathcal H}}{2} (w + 2) \beta \alpha^{k j} \alpha^{i n} {\mathcal Z}^{m}(\alpha) \epsilon_{k m n}\biggr\}.
\label{ES20}
\end{eqnarray}
The first line at the right hand side of Eq. (\ref{ES20}) does not contain any spatial gradient and it 
is therefore ${\mathcal O}(\alpha)$. The remaining part of the expression at the right hand side 
of the relation reported in Eq. (\ref{ES20}) are instead ${\mathcal O}(\beta)$. 
Sticking to the situation 
treated in the present paper the explicit form of $\beta_{ij}(\vec{x},\tau)$ can be determined in terms 
of $\alpha_{ij}(\vec{x})$ by solving the remaining Einstein equations written in terms of the ADM decomposition 
\cite{ADM1,ADM2}.  For this purpose Eqs.  (\ref{00}) and (\ref{ij})  can be written, respectively,  as 
\begin{eqnarray}
&&\partial_{\tau} K - N {\mathrm Tr} K^2  = \frac{N \ell^2_{\mathrm{P}}}{2}(3 p + \rho),
\label{FORM2}\\
&& \partial_{\tau} K_{i}^{j} - N K K_{i}^{j} - N r_{i}^{j} = 
\frac{N \ell_{\mathrm{P}}^2}{2} (p - \rho) \delta_{i}^{j}.
\label{FORM3}
\end{eqnarray}
Using Eqs. (\ref{ES3}) and (\ref{ES4}) into Eqs. (\ref{FORM2}), the following pair of conditions 
are obtained 
\begin{eqnarray}
&& \partial_{\tau}\biggl(\frac{\partial_{\tau} \beta}{2 a }\biggr) + \frac{{\mathcal H}}{a} \partial_{\tau} \beta = - 
\frac{a \ell_{\mathrm{P}}^2}{2} (3 p^{(1)} + \rho^{(1)}),
\label{SOL1}\\
&& \partial_{\tau} {\mathcal H} = - \frac{a^2 \ell_{\mathrm{P}}^2}{2} ( \rho^{(0)} + 3 p^{(0)}).
\label{SOL1a}
\end{eqnarray} 
To obtain Eqs. (\ref{SOL1}) and (\ref{SOL1a}) the total pressure and the total energy density 
have been separated as:
\begin{equation}
p(\vec{x},\tau) = p^{(0)}(\tau)  + p^{(1)}(\vec{x},\tau), \qquad \rho(\vec{x},\tau) = \rho^{(0)}(\tau) + \rho^{(1)}(\vec{x},\tau),
\label{SOL1b}
\end{equation}
where $p^{(1)}(\vec{x},\tau)$ and $\rho^{(1)}(\vec{x},\tau)$ vanish in the conformally flat limit.
Using Eqs. (\ref{ES3}) and (\ref{ES4}) into Eqs. (\ref{FORM3}) two further equations are obtained 
and they are:
\begin{eqnarray}
&& \partial_{\tau} \biggl( \frac{\partial_{\tau} \beta_{i}^{j}}{2 a} \biggr) + 
{\mathcal H} \frac{\partial_{\tau} \beta}{2 a} \delta_{i}^{j} + 
\frac{3 {\mathcal H}}{2 a } \partial_{\tau} \beta_{i}^{j} + a r_{i}^{j} = - \frac{a \ell_{\mathrm{P}}^2}{2} (p^{(1)} - \rho^{(1)}) \delta_{i}^{j},
\label{SOL2}\\
&& \partial_{\tau} {\mathcal H} + 2 {\mathcal H}^2 = - 
\frac{\ell_{\mathrm{P}}^2 a^2}{2}  (p^{(0)} - \rho^{(0)}).
\label{SOL2a}
\end{eqnarray}
Solving Eqs. (\ref{SOL1a}) and (\ref{SOL2a}) under the hypothesis of constant barotropic index (already 
assumed in Eq. (\ref{ES18})), $ p^{(1)}$ and  $\rho^{(1)}$ can be eliminated 
between Eqs. (\ref{SOL1}) and (\ref{SOL2}) and it turns out that $\beta_{ij}(\vec{x},\tau)$ obeys 
the following evolution equation:
\begin{equation}
\partial_{\tau}^2 \beta_{i}^{j} + 2 {\mathcal H} \partial_{\tau} \beta_{i}^{j} + 
\delta_{i}^{j} \biggl( \frac{1 - w}{1 + 3 w} \partial_{\tau}^2 \beta + 2 \frac{1+w}{1 + 3 w}
{\mathcal H} \partial_{\tau} \beta\biggr)  
 + 2 a^2 r_{i}^{j} =0.
\label{FORM3a}
\end{equation}
By solving Eq. (\ref{FORM3a}) the explicit form of $\beta_{ij}$ can be written in a separable form as 
$\beta_{i}^{j}(\vec{x},\tau) = g(\tau) \mu_{i}^{j}(\vec{x})$ where:
\begin{eqnarray}
&& g(\tau) = a^{3 w +1}, 
\label{FORM4}\\
&& \mu_{i}^{j}(\vec{x}) = - \frac{4}{H_{\mathrm{i}}^2 ( 3 w + 5) ( 3 w +1)} \biggl[ P_{i}^{j}(\vec{x}) 
+ \frac{3 w^2 - 6 w - 5}{4 ( 9 w + 5)} P(\vec{x}) \delta_{i}^{j} \biggr].
\label{FORM5}
\end{eqnarray}
Note that $P_{i}^{j}(\vec{x}) = r_{i}^{j}(\vec{x},\tau) a^2(\tau)$ accounts for the 
intrinsic curvature computed from $\alpha_{ij}(\vec{x})$. In Eqs. (\ref{FORM2}) and (\ref{FORM3}) the contribution 
of the velocity fields and of the magnetic fields 
has been neglected because they are subleading to ${\mathcal O}(\beta)$.
In the following two sections we will therefore present the full estimate 
of the vorticity to first-order in the gradient expansion. If needed the first-order 
result, together with Eqs. (\ref{FORM4}) and (\ref{FORM5}) can be used to estimate the 
vorticity to higher order. 

\renewcommand{\theequation}{5.\arabic{equation}}
\setcounter{equation}{0}
\section{Vorticity to first-order in the gradient expansion}
\label{sec5}
The simplest parametrization of $\alpha_{ij}(\vec{x})$ which does 
not contain spatial gradients can be written as 
\begin{equation}
\alpha_{ij}(\vec{x}) = e^{- 2 \Psi(\vec{x})} \delta_{ij}, \qquad \alpha = \mathrm{det}\,\alpha_{ij} = e^{- 6 \Psi(\vec{x})}.
\label{F0}
\end{equation}
In this case it is easy to show that ${\mathcal Z}^{m}(\alpha)= 0$ and therefore the first-order in the gradient expansion vanishes identically. In the $\Lambda$CDM scenario the scalar mode appearing in Eq. (\ref{F0}) 
leads to a $|\Psi(\vec{x})| \ll 1$ and therefore, in practice,  $\alpha_{ij}(\vec{x})$ is accurately 
estimated by  $\delta_{ij} - 2 \Psi(\vec{x}) \delta_{ij}$. To have a ${\mathcal Z}^{m}(\alpha) \neq 0$ 
the contribution of the 
tensor modes must be included and $\alpha_{ij}(\vec{x})$ will then given by:
\begin{equation}
\alpha_{ij}(\vec{x}) = \biggl[ \delta_{ij} + h_{ij}(\vec{x})\biggr],\qquad 
 \alpha^{ij}(\vec{x}) =  \biggl[ \delta^{ij} - h^{ij} + h^{i k} h_{k}^{j}\biggr],\qquad\sqrt{\alpha} = \biggl[ 1 - \frac{1}{4} h_{i}^{k} \, h_{k}^{i} \biggr],
\label{F0a}
\end{equation}
where $h_{ij}$ is divergenceless and traceless, i.e. $\partial_{i} h^{i j}= h_{i}^{i}= 0$. It must be borne in mind that 
the scalar and the tensor modes, in the $\Lambda$CDM scenario and in its tensor extension, are defined in terms 
of the conventional perturbative expansion. As a consequence of the latter statement, the informations 
on the spatial inhomogeneities of the model are not specified by assigning the analog 
$\alpha_{ij}( \vec{x})$ (or $\gamma_{ij}(\vec{x},\tau)$  to a given order in the spatial gradients). 
On the contrary, as it is more natural, the scalar and tensor modes of the geometry are 
specified by assigning the corresponding power spectra at a given pivot scale. To evaluate 
the appropriate correlators defining the vorticity we shall need first to obtain  
the fluctuations in real space (as opposed to Fourier space). Therefore, as we will show in the 
present and in the following section, the idea will be first to compute the fluctuations in real space 
and then to use the obtained result for the determination of the correlators defining the vorticity. 
This procedure will circumvent the calculation of complicated convolutions and will also be perfectly 
suitable for the applications described in section \ref{sec6}. 
Using then Eq. (\ref{ES13}) we have that
\begin{equation}
{\mathcal Z}^{m}(\alpha) = \partial_{q} \alpha^{q m} + \alpha^{m q} \alpha^{a b} \partial_{b} \alpha_{q a}
= h^{ q m} \, h^{a b}\, \partial_{b} h_{q a} + h^{a p}\, h_{p}^{b}\, \partial_{b} h_{a}^{m}.
\label{F0b}
\end{equation}
From Eq. (\ref{ES20}) the tensor ${\mathcal A}^{ij}(\alpha,\beta)$ can be computed to lowest order 
(i.e. by setting $\beta=0$) and the result will therefore be written, using Eq. (\ref{F0b}), as
\begin{equation}
{\mathcal A}^{ij}(\alpha) =- \frac{{\mathcal H}}{3 {\mathcal H}_{1}^2 (w + 1)} \biggl(\frac{a}{a_1}\biggr)^{3 w +1} \, \epsilon^{m i j} 
\biggl[ h^{a \ell} h_{\ell}^{b} \partial_{b} h_{a m} + h_{m q} h^{b a } \partial_{b} h^{q}_{a} \biggr]  
 + {\mathcal O}(\epsilon^2).
\label{F2}
\end{equation}
Finally, the total vorticity can be derived directly from Eq. (\ref{ES19}) 
\begin{eqnarray}
\omega^{i}_{\mathrm{tot}} &=& - {\mathcal L}(\tau,w) \, \epsilon^{m i j} \partial_{j}\biggl[ h^{a \ell} h_{\ell}^{b} \partial_{b} h_{a m} + h_{m q} h^{b a } \partial_{b} h_{q a} \biggr]  + {\mathcal O}(\epsilon^3),
\nonumber\\
{\mathcal L}(\tau,w) &=& \frac{{\mathcal H}}{3 {\mathcal H}_{1}^2 (w + 1)} \biggl(\frac{a}{a_1}\biggr)^{3 w +1}.
\label{F3}
\end{eqnarray}
To give an explicit estimate of the primordial vorticity  the relevant cosmological parameters will be taken to be the 
ones determined on the basis of the WMAP 7yr data alone \cite{wmap7a,wmap7b}.
In the $\Lambda$CDM paradigm the sole source of curvature inhomogeneities is represented by the 
standard adiabatic mode whose associated power spectrum is assigned 
at the comoving pivot scale $k_{\mathrm{p}} = 0.002\, \mathrm{Mpc}^{-1}$
with characteristic amplitude ${\mathcal A}_{{\mathcal R}}$
\begin{equation}
\langle {\mathcal R}(\vec{k},\tau) {\mathcal R}(\vec{p},\tau) \rangle = \frac{2 \pi^2}{k^3} 
{\mathcal P}_{{\mathcal R}}(k) \delta^{(3)}(\vec{k} + \vec{p}), \qquad {\mathcal P}_{{\mathcal R}}(k) = {\mathcal A}_{{\mathcal R}} \biggl(\frac{k}{k_{\mathrm{p}}}\biggr)^{n_{\mathrm{s}}-1},
\label{AD1}
\end{equation} 
where $n_{\mathrm{s}}$ denotes the spectral index associated with the fluctuations of the spatial curvature. According to the WMAP 7yr data alone analyzed in the light of the $\Lambda$CDM paradigm and without tensors modes \cite{wmap7a,wmap7b} the determinations 
of ${\mathcal A}_{{\mathcal R}}$ and of $n_{\mathrm{s}}$ lead, respectively, to 
${\mathcal A}_{{\mathcal R}} = (2.43 \pm 0.11)\times 10^{-10} $ and to $n_{\mathrm{s}} = 0.963 \pm 0.014$.
The standard $\Lambda$CDM scenario, sometimes dubbed vanilla $\Lambda$CDM is defined by six pivotal parameters 
whose specific values are, in the absence of tensor modes\footnote{Following the standard notations (slightly modified to 
avoid possible clashes with previously defined variables) 
$\Omega_{\mathrm{b}}, \, \Omega_{\mathrm{c}}, \Omega_{\mathrm{de}}$ denote, respectively, the present critical fractions 
 of thebaryons, of the dark matter, of the dark energy; $h_{0}$ is the Hubble constant in units of $100\, \mathrm{km}/(\mathrm{sec}\, \mathrm{Mpc})$, $n_{\mathrm{s}}$ is the scalar spectral index while $\epsilon_{\mathrm{re}}$ denotes 
 the optical depth at recombination.}
\begin{equation}
( \Omega_{\mathrm{b}}, \, \Omega_{\mathrm{c}}, \Omega_{\mathrm{de}},\, h_{0},\,n_{\mathrm{s}},\, \epsilon_{\mathrm{re}}) \equiv 
(0.0449,\, 0.222,\, 0.734,\,0.710,\, 0.963,\,0.088).
\label{PP1}
\end{equation}
To estimate the correlation functions associated with Eqs. (\ref{F2}) and (\ref{F3}) 
it is mandatory to know in detail the numerical value 
of the correlation function of the tensor modes of the geometry which have not been  
detected so far but whose specific upper limits will determine the maximal magnetic field 
obtainable from the vorticity of the geometry.  The tensor modes of the geometry 
are described in terms of a rotationally and parity invariant two-point function 
\begin{equation}
\langle h_{ij}(\vec{x},\tau) \, h_{ij}(\vec{y},\tau) \rangle = \int \frac{dk}{k} {\mathcal P}_{\mathrm{T}}(k,\tau) \, \frac{\sin{k r}}{kr},\qquad 
\label{PP2}
\end{equation}
where the tensor power spectrum at the  generic time $\tau$ is given by the product of the appropriate transfer 
function multiplied by the primordial spectrum:
\begin{equation}
{\mathcal P}_{\mathrm{T}}(k,\tau) = {\mathcal M}(k,\, k_{\mathrm{eq}},\, \tau) \overline{{\mathcal P}}_{\mathrm{T}}(k), \qquad 
\overline{{\mathcal P}}_{\mathrm{T}}(k) = {\mathcal A}_{\mathrm{T}} \biggl(\frac{k}{k_{\mathrm{p}}}\biggr)^{n_{\mathrm{T}}};
\label{PP2a}
\end{equation}
note that ${\mathcal A}_{\mathrm{T}}$ is the amplitude of the tensor power spectrum and $n_{\mathrm{T}}$ is the tensor 
spectral index. The transfer function ${\mathcal M}(k,\, k_{\mathrm{eq}},\, \tau)$ can be computed under several approximations depending 
upon the required accuracy. The transfer function for the amplitude of the tensor modes can be numerically 
computed by solving the evolution of the tensor fluctuations across the matter-radiation equality and the result is \cite{mgt1,mgt2}
\begin{equation} 
 {\mathcal M}(k,\, k_{\mathrm{eq}},\, \tau)= \frac{9\,j^2_{1}(k\tau)}{|k\tau|^2}
 \biggl[ 1 + c_{1} \biggl(\frac{k}{k_{\mathrm{eq}}}\biggr) + c_{2}\biggl(\frac{k}{k_{\mathrm{eq}}}\biggr)^2\biggr], 
\label{PP2b}
\end{equation}
where\footnote{The analysis of \cite{tt1} gave $c_{1}= 1.34$ and $c_{2}= 2.50$ 
which is fully compatible with the results of \cite{mgt1,mgt2}. In the approach of \cite{tt1} (see also \cite{tur1}) the calculation of the amplitude transfer function, in fact, involve a delicate matching on the phases of the tensor mode 
functions. Conversely, if the transfer function is computed directly for the spectral energy density, the oscillatory contributions are suppressed as the wavelengths get shorter than the Hubble radius (see below).},
 according to \cite{mgt1,mgt2}, $c_{1} = 1.26$ and $c_{2} = 2.68$.
In Eq. (\ref{PP2b}) $j_{1}(y) = (\sin{y}/y^2 - \cos{y}/y)$ is the spherical Bessel function of first kind which is related to the approximate solution of the evolution equations for the tensor mode functions whenever the solutions  are computed deep in the matter-dominated phase (i.e. $a(\tau) \simeq \tau^2$). Instead of working directly with ${\mathcal A}_{\mathrm{T}}$ it 
is often preferred to introduce the quantity customarily called $r_{\mathrm{T}}$ denoting  the 
ratio between the tensor and the scalar amplitude at the pivot scale $k_{\mathrm{p}}$ 
\begin{equation}
r_{\mathrm{T}} = \frac{{\mathcal A}_{\mathrm{T}}}{{\mathcal A}_{{\mathcal R}}}=  \frac{\overline{{\mathcal P}}_{\mathrm{T}}(k_{\mathrm{p}})}{{\mathcal P}_{{\mathcal R}}(k_{\mathrm{p}})}.
\label{PP3}
\end{equation}
In principle $n_{\mathrm{T}}$ can be taken to be independent of $r_{\mathrm{T}}$ and this possibility will also 
be contemplated in the present discussion. At the same time, if the scalar and the tensor modes 
are both of inflationary origin, $n_{\mathrm{T}}$ is related to $r_{\mathrm{T}}$ and to the slow-roll
parameter $\epsilon$ which measure the rate of decrease of the Hubble parameter 
during the conventional inflationary stage of expansion:
\begin{equation}
n_{\mathrm{T}} = - \frac{r_{\mathrm{T}}}{8} = - 2 \epsilon, \qquad \epsilon = - \frac{\dot{H}}{H^2};
\label{PP4}
\end{equation}
 the overdot denotes the usual  derivative with respect to the cosmic time coordinate; in Eq. (\ref{PP4}) 
the spectral index is frequency-independent but there exist situations where more general possibilities 
can be contemplated  such as, for instance
\begin{equation}
n_{\mathrm{T}} = - 2 \epsilon + \frac{\alpha_{\mathrm{T}}}{2} \ln{(k/k_{\mathrm{p}})}, \qquad \alpha_{\mathrm{T}} = \frac{r_{\mathrm{T}}}{8}\biggl[ (n_{\mathrm{s}} -1) + \frac{r_{\mathrm{T}}}{8}\biggr]. 
\label{PP5}
\end{equation}
If $\alpha_{\mathrm{T}} =0$ the tensor spectral index $n_{\mathrm{T}}$ does not depend upon the frequency and this is the case which is, somehow, endorsed when introducing gravitational waves in the minimal tensor extension of the $\Lambda$CDM.  If a tensor component is allowed in the analysis 
of the WMAP 7yr data alone the relevant cosmological parameters are determined to be 
\begin{equation}
( \Omega_{\mathrm{b}}, \, \Omega_{\mathrm{c}}, \Omega_{\mathrm{de}},\, h_{0},\,n_{\mathrm{s}},\, \epsilon_{\mathrm{re}}) \equiv 
(0.0430,\, 0.200,\, 0.757,\,0.735,\, 0.982,\,0.091).
\label{Par2}
\end{equation}
In the case of Eq. (\ref{PP1}) the amplitude of the scalar modes is ${\mathcal A}_{{\mathcal R}} = 
(2.43 \pm 0.11) \times 10^{-9}$ while in the case of Eq. (\ref{Par2}) the corresponding values of ${\mathcal A}_{{\mathcal R}}$ and of $r_{\mathrm{T}}$ are given by 
\begin{equation}
{\mathcal A}_{{\mathcal R}} = (2.28 \pm 0.15)\times 10^{-9},\qquad r_{\mathrm{T}} < 0.36,
\label{Par3}
\end{equation}
to $95$ \% confidence level. To avoid confusions it is appropriate 
to spend a word of care on the figures implied by Eqs. (\ref{Par2}) and 
(\ref{Par3}) which have been used in the numeric analysis just 
for sake of accuracy. The qualitative features 
of the effects discussed here do not change if, for instance, one 
would endorse the parameters drawn from the comparison of the minimal tensor extension 
of the $\Lambda$CDM with the WMAP 5yr data release \cite{wmap5a,wmap5b}, implying, for instance, ${\mathcal A}_{{\mathcal R}} = 2.1^{+2.2}_{-2.3}\times 10^{-9}$, $n_{\mathrm{s}} =0.984$ and  $r_{\mathrm{T}} < 0.65$ (95 \% confidence level). Similar orders 
of magnitude can be also obtained from even older releases \cite{wmap3,wmapfirst}.

\renewcommand{\theequation}{6.\arabic{equation}}
\setcounter{equation}{0}
\section{Magnetic field induced by the total vorticity}
\label{sec6}
The total vorticity derived in the previous sections is larger than the vorticity of the ions. Therefore,  the total magnetic field derived on the basis of $\omega_{\mathrm{tot}}^{i}$ is larger than the one derived on the basis of the ion contribution. Of course this statement holds in an averaged sense since what matters 
is not the vorticity itself but rather its two-point function which will be explicitly 
computed in the present section. Using Eq. (\ref{F3}) the maximal obtainable magnetic field 
will be the one given by Eqs. (\ref{D7})--(\ref{D8}) or (\ref{D12}) where the total vorticity induced by the geometry is given by Eq. (\ref{F3}) 
\begin{equation}
B_{\mathrm{max}}^{i}(\vec{x},\tau) = - \frac{\rho_{\mathrm{i}}  \sqrt{\gamma} }{ e\,N^2 \tilde{n}_{\mathrm{i}}} \omega^{i}_{\mathrm{tot}}(\vec{x},\tau).
\label{Fmax1}
\end{equation}
which can also be written, by explicitly keeping track of the number of gradients, as  
\begin{equation} 
B_{\mathrm{max}}^{i}(\vec{x},\tau) = \biggl\{{\mathcal L}(\tau,w)  \, \epsilon^{m i j} \partial_{j}\biggl[ h^{a \ell} h_{\ell}^{b} \partial_{b} h_{a m} + h_{m q} h^{b a } \partial_{b} h^{q}_{a} \biggr] + {\mathcal O}(\epsilon^3) \biggr\} \, a(\tau)\, \biggl[ 1 + {\mathcal O}(\epsilon^2) \biggr].
\label{Fmax2}
\end{equation}
The prefactor appearing in Eq. (\ref{Fmax1}) has been estimated, in Eq. (\ref{Fmax2}),  by recalling 
that, to lowest order in the gradient expansion
\begin{equation}
\partial_{\tau} \rho_{\mathrm{i}} = N K \rho_{\mathrm{i}}, \qquad \partial_{\tau} \tilde{n}_{\mathrm{i}} = N K \tilde{n}_{\mathrm{i}},
\label{pref1}
\end{equation}
implying that $\rho_{\mathrm{i}}$ and $\tilde{n}_{\mathrm{i}}$ scale in the same way with $\sqrt{\gamma}$ since 
$ N K = - \partial_{\tau} \ln{\sqrt{\gamma}}$. But then, from Eqs. (\ref{FORM4}), (\ref{FORM5}) and (\ref{F0a}):
\begin{equation}
 \frac{\rho_{\mathrm{i}}  \sqrt{\gamma} }{ N^2 \tilde{n}_{\mathrm{i}}} = a(\tau) \biggl[ 1 + {\mathcal O}(\epsilon^2)\biggr],
\label{pref2}
\end{equation}
where the first correction is ${\mathcal O}(\epsilon^2)$ and depends on $\beta$ (see, e.g. Eqs. (\ref{FORM4}) and (\ref{FORM5})) but it will be immaterial for the present ends. 
From now on the subscripts will be dropped but it will be always understood that we are referring here to the total vorticity and to the maximally achievable magnetic field. As a consequence of Eq. (\ref{Fmax1}) the correlation function of the magnetic field can be related to the correlation function of the vorticity. 
To estimate the correlation of the vorticity  and to obtain an explicit expression 
the key point is to reduce the six-point function of the 
tensor modes to the product of two-point functions. For this purpose it is not sufficient to consider 
the trace of the two-point function introduced in Eq. (\ref{PP2}) but it is rather necessary 
to proceed with the full tensorial structure of the correlator whose general parity and rotationally-invariant form will be denoted as
\begin{equation}
G_{i j m n}(r) = \langle h_{ij}(\vec{x},\tau) \, h_{m n}(\vec{y},\tau) \rangle,
\label{F6}
\end{equation}
where $G_{i j m n}(r)$ is only function of $r = |\vec{r}|$ where $\vec{r} = \vec{x} - \vec{y}$. Since both 
$h_{i j}(\vec{x},\tau)$ and $h_{m n}(\vec{y},\tau)$ are transverse and traceless,  $G_{i j m n}(r)$ will have to share 
the same properties. In particular, $G_{i j m n}(r)$ must be symmetric for $i\to j$, $m \to n$, $(i\,j)\to(m\, n)$ and satisfy 
the following properties  
\begin{eqnarray}
&& \frac{\partial}{\partial r^{i}} G_{i j m n}=0, \qquad G_{i i m n} = G_{i j m m} = 0 
\label{F7}\\
&& \mathrm{Tr}[ G_{i j m n} ] = G_{i j i j}= \int \frac{dk}{k} {\mathcal P}_{\mathrm{T}}(k) \, \frac{\sin{k r}}{kr}.
\label{F8}
\end{eqnarray}
The properties of Eq. (\ref{F7}) and (\ref{F8}) are a reflection of the divergenceless and traceless 
nature of $h_{ij}(\vec{x},\tau)$ while the requirement on the trace follows from the consistency with Eq. (\ref{PP2}).
The general form of $G_{i j m n}$ can therefore be written as 
\begin{eqnarray}
G_{i j m n}(r) &=& \biggl(\delta_{i m} \delta_{n j} + \delta_{m j} \delta_{n i} \biggr)\, G_{1}(r) + \delta_{i j} \delta_{m n}\, G_{2}(r)
\nonumber\\
&+& \biggl(\delta_{i j} \, r_{m}\, r_{n}\, + \delta_{m n}\, r_{i}\, r_{j}\biggr) G_{3}(r) 
\nonumber\\
&+& \biggl(\delta_{j n} r_{i} r_{m} + 
\delta_{i m} r_{j} r_{n}  + \delta_{j m} r_{i} r_{n} + \delta_{i n} r_{j} r_{m}\biggr) G_{4}(r) 
\nonumber\\
&+&  r_{i}\, r_{j}\, r_{m}\,r_{n} G_{5}(r), 
\label{F9A}
\end{eqnarray}
where the various independent functions appearing in Eq. (\ref{F9A}) are determined in appendix \ref{APPB}.
The methods used to analyze the real-space correlators are the ones exploited in usual 
applications of statistical fluid mechanics \cite{monin,stoc}. 
To evaluate Eq. (\ref{Fmax2}) can then proceed as follows. Using Eq. (\ref{F3}), the explicit form of the correlator 
of the vorticity becomes
\begin{eqnarray}
 \langle \omega^{i}(\vec{x},\tau)\, \omega^{i}(\vec{y},\tau) \rangle &=& {\mathcal L}^2(\tau,w) \epsilon^{j m i}\, \epsilon^{j' m' i} \frac{\partial^2}{\partial y^{j'}\, \partial y^{b'}} \frac{\partial^2 }{\partial x^{j} \, \partial x^{b}} {\mathcal T}_{b m b' m'}
 \nonumber\\
  {\mathcal T}_{b m b' m'} &=&
 \langle\biggl[ h_{a\, b} \, h_{ a \, q} \, h_{q \,m}\biggr]_{\vec{x}} \, \biggl[ h_{a'\, b'} h_{a' \, q'} h_{q' \, m'}\biggr]_{\vec{y}} \rangle. 
\label{F12}
\end{eqnarray}
By defining $\langle \omega^2(r) \rangle =  \langle \omega^{i}(\vec{x},\tau)\, \omega^{i}(\vec{y},\tau) \rangle$ and by
 recalling the notations of appendix \ref{APPB},  we shall have that\footnote{Recall that $\vec{r} = \vec{x} - \vec{y}$, i.e. $ r= |\vec{r}| = | \vec{x} - \vec{y}|$.}  
\begin{equation}
\langle\omega^2(r)\rangle = {\mathcal L}^2(\tau,w) \epsilon^{j m i}\, \epsilon^{j' m' i} \frac{\partial^2}{\partial r^{j'}\, \partial r^{b'}}
\frac{\partial^2 }{\partial r^{j} \, \partial r^{b}} {\mathcal T}_{m' b' m b},
\label{F13}
\end{equation}
where the quantity ${\mathcal T}_{m' b' m b}$ is a function of $r$; the explicit form of  ${\mathcal T}_{m' b' m b}$ is given in
appendix \ref{APPB} in terms of the two-point functions $G_{i j m n}$. Furthermore, since 
$T_{m' b' m b}$ can be written, in general terms, as
\begin{eqnarray}
{\mathcal T}_{m' b' m b} &=& T_{1}(r) (\delta_{m' b} \delta_{m b'} + \delta_{m m'} \delta_{b b'})
+ T_{2}(r) \delta_{m' b'} \delta_{m b}
\nonumber\\
&+& T_{3}(r) ( \delta_{m' b'} r_{m} r_{b} + \delta_{m b} r_{m'} r_{b'}) + T_{4}(r) ( \delta_{b m'} r_{m} r_{b'} + 
\delta_{m b'} r_{b} r_{m'}  
\nonumber\\
&+& \delta_{m m'} r_{b} r_{b'} + \delta_{b b'} r_{m} r_{m'}) + r_{m'} r_{b'} r_{m} r_{b} T_{5}(r).
\label{F15}
\end{eqnarray}
By using the results of appendix \ref{APPB} the explicit values of the five $T_{i}(r)$ can be expressed in terms 
of the two-point functions $G_{i j m n}$ (see Eq. (\ref{AB2}) and (\ref{AB3})--(\ref{AB7})). 
There are two physically complementary regimes where the primordial vorticity and hence 
the magnetic field can be evaluated. Comoving lengths $r_{\mathrm{G}}$ defined between $1$ and $100$ Mpc are smaller than the Hubble radius at equality since\footnote{The quantity $r_{\mathrm{T}}$ (denoting, in sec. \ref{sec5}, the tensor 
to scalar ratio) must not be confused with $r_{\mathrm{G}}$ and $r_{\mathrm{eq}}$ which denote specific 
values of the radial coordinate.}
\begin{equation}
r_{\mathrm{eq}} = 
\frac{2 (\sqrt{2} -1)}{H_{0}} \frac{\sqrt{ \Omega_{\mathrm{R}0}}}{ \Omega_{\mathrm{M}0}} 
= 119.397 \biggl(\frac{h_{0}^2 \Omega_{\mathrm{M}0}}{0.134}\biggr)^{-1} \biggl(\frac{h_{0}^2 \Omega_{\mathrm{R}0}}{4.15 \times 10^{-5}}\biggr)^{1/2}\,\,\mathrm{Mpc},
\label{TA13}
\end{equation}
where $H_{0}$ is the present value of the Hubble rate, $\Omega_{\mathrm{M}0}$ is the 
present value of the critical fraction in matter and $\Omega_{\mathrm{R}0}$ is the 
present value of the critical fraction in radiation. The pivot length $r_{\mathrm{p}} = 500\, \mathrm{Mpc}$ 
at which the tensor amplitudes are assigned is such that $r_{\mathrm{G}} < r_{\mathrm{eq}} < r_{\mathrm{p}
}$. Therefore, after matter-radiation equality and, in particular, at photon decoupling, the correlation function 
of the magnetic field can be estimated as
\begin{eqnarray}
\langle B^2(r) \rangle &=& 6.348 \times 10^{-76} \biggl(\frac{r_{\mathrm{T}}}{0.32}\biggr)^3 \biggl(\frac{{\mathcal A}_{{\mathcal R}}}{2.43 \times 10^{-9}}\biggr)^{3} \biggl(\frac{z_{\mathrm dec} +1}{1089.2}\biggr)^2 
\nonumber\\
&\times& \biggl(\frac{h_{0}^2 \Omega_{\mathrm{M}0}}{0.134}\biggr)^{6}\, 
\biggl(\frac{h_{0}^2 \Omega_{\mathrm{R}0}}{4.15 \times 10^{-5}}\biggr)^{-6}\,\,  {\mathcal C}(n_{\mathrm{T}}, r)
\,\, \mathrm{G}^2,
\nonumber\\
 {\mathcal C}(n_{\mathrm{T}}, r) &=&  c(n_{\mathrm{T}}) \biggl(\frac{r}{r_{\mathrm{p}}}\biggr)^{ 8 - 3 n_{\mathrm{T}}}
+ d(n_{\mathrm{T}}),
\label{TA14}
\end{eqnarray}
in units of $\mathrm{G}^2\equiv \mathrm{Gauss}^2$ and 
where the constants $c(n_{\mathrm{T}})$ and $d(n_{\mathrm{T}})$ are given by\footnote{Recall that because of the relation (\ref{PP4}) $n_{\mathrm{T}} < 0$ and $r_{\mathrm{T}}>0$.} 
\begin{eqnarray}
c(n_{\mathrm{T}}) &=& - 2 (n_{\mathrm{T}} - 4 ) (n_{\mathrm{T}} - 3 )[ 2 n_{\mathrm{T}}(n_{\mathrm{T}} - 6) + 19]
\cos^3{\biggl(\frac{n_{\mathrm{T}} \pi}{2}\biggr)} \Gamma^3(n_{\mathrm{T}} - 5),
\nonumber\\
d(n_{\mathrm{T}}) &=& - \frac{36 + 14 n_{\mathrm{T}} (n_{\mathrm{T}} - 4)}{ 45 n_{\mathrm{T}} ( n_{\mathrm{T}}^2 - 6n_{\mathrm{T}} + 8 )^2}.
\label{TA15}
\end{eqnarray}
The typical values of $n_{\mathrm{T}}$ are negative and ${\mathcal O}(10^{-2})$. Indeed, 
assume, consistently with Eq. (\ref{Par3}), that $r_{\mathrm{T}} \sim 0.32$. Then, according to Eq.  (\ref{PP4}), $n_{\mathrm{T}} \sim - 0.04$ 
and $\epsilon \sim 0.02$.
Concerning the results of Eqs. (\ref{TA14}) and (\ref{TA15}) few comments are in order:
\begin{itemize}
\item{} the prefactor ${\mathcal L}(\tau,w)$ is estimated in the hypothesis 
$w=0$, $a_{1} = a_{\mathrm{eq}}$ and ${\mathcal H}_{1} = {\mathcal H}_{\mathrm{eq}}$ since 
we ought to estimate the field prior to photon decoupling;
\item{} recalling that ${\mathcal H} = a H$ the value of the Hubble rate at the equality time can be 
estimated as:
\begin{equation}
H_{\mathrm{eq}} =  \sqrt{ 2 \,\,\Omega_{\mathrm{M}0}} \, H_{0} \, \biggl(\frac{a_{0}}{a_{\mathrm{eq}}}\biggr)^{3/2} \equiv 1.65 \times 10^{-56} \biggl(\frac{h_{0}^2 \Omega_{\mathrm{M}0}}{0.134}\biggr)^2\, M_{\mathrm{P}};
\end{equation}
\item{} the result of Eq. (\ref{TA14}) holds for comoving scales $r < r_{\mathrm{P}} = 500$ Mpc 
(which are the ones relevant for the gravitational collapse of the protogalaxy) and it is not sensitive to the variation of $r$ provided $n_{\mathrm{T}}$ is nearly scale-invariant;
\item{} if $r_{\mathrm{T}} \simeq 0.32$, then $n_{\mathrm{T}} = - 0.04$; from Eq. (\ref{TA15}), ${\mathcal C}(r_{\mathrm{G}}, -0.04) \simeq 0.07$ while for $r= 100\, r_{\mathrm{G}}$ we have 
 that  ${\mathcal C}(100\, r_{\mathrm{G}}, -0.04) \simeq 0.01$.
\end{itemize}
By thus approximating ${\mathcal C}(n_{\mathrm{T}}, r) \simeq {\mathcal O}(1)$  
 in the range  $r= 1$--$100$ Mpc and for 
$0.2 < r_{\mathrm{T}} < 0.3$ we get the following value for $B_{\mathrm{max}}=  \sqrt{\langle B^2(r) \rangle}$ 
\begin{eqnarray} 
B_{\mathrm{max}} &=&  2.519 \times 10^{-38} \biggl(\frac{r_{\mathrm{T}}}{0.32}\biggr)^{3/2} \biggl(\frac{{\mathcal A}_{{\mathcal R}}}{2.43 \times 10^{-9}}\biggr)^{3/2} \biggl(\frac{z_{\mathrm dec} +1}{1089.2}\biggr) 
\nonumber\\
&\times& \biggl(\frac{h_{0}^2 \Omega_{\mathrm{M}0}}{0.134}\biggr)^{3}\, 
\biggl(\frac{h_{0}^2 \Omega_{\mathrm{R}0}}{4.15 \times 10^{-5}}\biggr)^{-3}\,\,\mathrm{G}.
\label{TA16}
\end{eqnarray}
The result of Eq. (\ref{TA16}) does not seem to be even remotely relevant for galactic magnetogenesis 
or for cluster magnetogenesis. In spite of the intricacy and of the ramification of the 
galactic dynamo hypothesis, it is useful to compare Eq. (\ref{TA16}) with the minimal 
requirements stemming from what we would call optimal or ideal dynamo, namely 
a process where the kinetic energy of the protrogalaxy is converted into magnetic energy with 
maximal efficiency. Let us denote with  $N_{\mathrm{rot}}$ the number of (effective) rotations performed 
 by the galaxy since gravitational collapse and with $\rho_{\mathrm{a}}$ and $\rho_{\mathrm{b}}$ 
the matter density after and before gravitational collapse   .
 
The typical rotation period of a spiral galaxy is of the order of $3\times10^{8}$ yrs which should be compared with $10^{10}$ yrs, i.e. the approximate age of the galaxy.  The maximal number of rotations 
 performed by the galaxy since its origin is then of the order of $N_{\mathrm{rot}}\sim 30$.
Under the hypothesis that the kinetic energy of the plasma 
 is transferred to the magnetic energy with maximal efficiency, the protogalactic field 
 will be amplified by one efold during each rotation. The effective 
 number of efolds is however always smaller than $30$ for various reasons. Typically it can happen that the dynamo quenches prematurely because some of the higher wavenumbers  
 of the magnetic field become critical (i.e. comparable with the kinetic energy of the plasma) before the smaller ones. Other sources of quenching have been recently discussed in the literature (see, for an introduction to this topic, section 4.2 of \cite{dyn} and references therein). There is also another source of amplification of the primordial magnetic field and it has to do with compressional 
amplification. At the time of the gravitational collapse of the 
protogalaxy the conductivity of the plasma was sufficiently high 
to justify the neglect of nonlinear corrections in the equations 
expressing the conservation of the magnetic flux and of  the 
magnetic helicity. The conservation of the magnetic flux
implies that, during the gravitational collapse, the magnetic field 
should undergo compressional amplification, i.e. the same 
kind of mechanism which is believed to be the source of the 
large magnetic fields of the pulsars. Taking into account the two previous 
observations the estimate of Eq. (\ref{TA16}) must be compared with the bound 
\begin{equation}
B_{\mathrm{bound}} \simeq 3 \times10^{3}\, e^{- N_{\mathrm{rot}}} \biggl(\frac{\rho_{\mathrm{b}}}{\rho_{\mathrm{a}}}\biggr)^{2/3} \,\, \mathrm{nG}
\label{TA17}
\end{equation}
in $\mathrm{nG}$ units.  Even assuming $N_{\mathrm{rot}} = 30$,  $\rho_{\mathrm{a}} \simeq 10^{-24} \, \mathrm{g}/\mathrm{cm}^3$, and $\rho_{\mathrm{b}} \simeq 10^{-29} \, \mathrm{g}/\mathrm{cm}^3$ the minimal value of $B_{\mathrm{bound}}$  is ${\mathcal O}(10^{-25}) \mathrm{G}$. Clearly, by comparing Eq. (\ref{TA16}) 
with Eq. (\ref{TA17}), $B_{\mathrm{max}} \ll B_{\mathrm{bound}}$.

Going then to cluster magnetogenesis, the typical scale of the gravitational collapse of a cluster is larger (roughly by one order of magnitude) than the scale of gravitational collapse of the protogalaxy. The mean mass density 
within the Abell radius ( $\simeq 1.5 h_{0}^{-1} $ Mpc) is roughly 
$10^{3}$ times larger than the critical density since  clusters are 
formed from peaks in the density field. Moreover, clusters 
rotate much less than galaxies even if it is somehow 
hard to disentangle, observationally, the global (coherent) 
rotation of the cluster from the rotation curves of the 
constituent galaxies. By assuming, for instance, $N_{\mathrm{rot}}=5$, a density gradient of $10^{3}$ and $500$ nG as final field, Eq. (\ref{TA17}) demands and initial seed of the order $0.15$ nG. 

Another application of the results obtained in the previous sections can be the estimate 
of the magnetic field induced by the total vorticity for scales which are 
larger than the Hubble radius prior to matter radiation equality. To conduct this estimate 
the explicit form of the correlators will change. First of all in the pre-factor 
${\mathcal L}(\tau, w)$ we shall choose $ w= 1/3$ and $a_{1} = a_{\mathrm{r}}$ and 
${\mathcal H}_{1} = {\mathcal H}_{\mathrm{r}}$ with $H_{\mathrm{r}} \simeq 10^{-5} \, M_{\mathrm{P}}$.
Thus for typical length-scales larger than the Hubble radius at equality and for typical times 
of the order of the equality time the analog of Eq. (\ref{TA14}) can be written as
\begin{eqnarray}
\langle B^2(r) \rangle &=& 2.915 \times 10^{-79} \biggl(\frac{r_{\mathrm{T}}}{0.32}\biggr)^3 \biggl(\frac{{\mathcal A}_{{\mathcal R}}}{2.43 \times 10^{-9}}\biggr)^{3} \biggl(\frac{z_{\mathrm dec} +1}{1089.2}\biggr)^2 
\nonumber\\
&\times& \biggl(\frac{h_{0}^2 \Omega_{\mathrm{M}0}}{0.134}\biggr)^{-4}\,  {\mathcal C}(n_{\mathrm{T}}, r)
\,\, \mathrm{G}^2,
\nonumber\\
 {\mathcal C}(n_{\mathrm{T}}, r) &=&  \tilde{c}(n_{\mathrm{T}}) \biggl(\frac{r}{r_{\mathrm{p}}}\biggr)^{ -4 - 3 n_{\mathrm{T}}}
+ \tilde{d}(n_{\mathrm{T}}),
\label{TA18}
\end{eqnarray}
where the numerical constants $\tilde{c}(n_{\mathrm{T}})$ and $\tilde{d}(n_{\mathrm{T}})$ are given by 
\begin{eqnarray}
\tilde{c}(n_{\mathrm{T}}) &=& - 2 n_{\mathrm{T}} ( n_{\mathrm{T}} + 1) [ 2 n_{\mathrm{T}}(n_{\mathrm{T}} + 2) + 3] 
\cos^3{\biggl(\frac{n_{\mathrm{T}} \pi}{2}\biggr)}  \Gamma^3(n_{\mathrm{T}} -1), 
\nonumber\\
\tilde{d}(n_{\mathrm{T}}) &=& - \frac{2( 7 n_{\mathrm{T}}^2 + 28 n_{\mathrm{T}} + 18)}{45 n_{\mathrm{T}}^2 (n_{\mathrm{T}}+ 2)^2 (n_{\mathrm{T}} + 4) }.
\label{TA19}
\end{eqnarray}
Equation (\ref{TA18}) holds under the assumption $r < r_{\mathrm{p}}$ which means, in practice, that 
it applies only for a narrow range of scales $ 120 \, \mathrm{Mpc} < r < 500 \, \mathrm{Mpc}$. If 
$r \simeq 250 \, \mathrm{Mpc}$ then $r/r_{\mathrm{p}} = 0.5$ and ${\mathcal C}(n_{\mathrm{T}}, r) \simeq 
{\mathcal O}(163)$ and 
\begin{eqnarray}
B_{\mathrm{max}} &=&  4.3 \times 10^{-39} \biggl(\frac{r_{\mathrm{T}}}{0.32}\biggr)^{3/2} \biggl(\frac{{\mathcal A}_{{\mathcal R}}}{2.43 \times 10^{-9}}\biggr)^{3/2}  \biggl(\frac{h_{0}^2 \Omega_{\mathrm{M}0}}{0.134}\biggr)^{-2}\, 
\,\,\mathrm{G}.
\label{TA20}
\end{eqnarray}

\renewcommand{\theequation}{7.\arabic{equation}}
\setcounter{equation}{0}
\section{Concluding remarks}
\label{sec7}
The idea explored in this paper has been  to compute the vorticity by employing a recently devised 
framework for the treatment of fully inhomogeneous plasmas which are 
also gravitating. The latter description brings a new perspective to the study of the evolution 
of the vorticity exchange in the electron-ion-photon system without 
postulating the customary separation between a (preferably 
conformally flat) background geometry and its relativistic fluctuations.
A set of general conservation laws has been derived on the 
basis of the fully inhomogeneous equations in different 
temperature regimes depending on the hierarchies between 
the exchange rate of the vorticity between electrons, ions and photons.
After expanding the Einstein equations as well as the vorticity 
equations to a given order in the spatial gradients, the total vorticity 
has then been estimated to lowest order in the gradient expansion. 

The maximal comoving magnetic field induced in the $\Lambda$CDM paradigm 
depends upon the tensor to scalar ratio and it is, at most, of the order 
of $10^{-37}$ G over the typical comoving scales ranging between $1$ and 
$10$ Mpc. The obtained results are irrelevant for seeding a reasonable
galactic dynamo action and they demonstrate how the proposed fully inhomogeneous treatment 
can be used for a systematic scrutiny of pre-decoupling plasmas beyond the conventional 
perturbative expansions. The estimate of the primordial vorticity induced in the 
$\Lambda$CDM scenario can also turn out to be relevant in related contexts 
such as the ones contemplated by non conventional paradigms of galaxy formation.

\newpage
\begin{appendix}
\renewcommand{\theequation}{A.\arabic{equation}}
\setcounter{equation}{0}
\section{Gradient expansion and pre-decoupling physics}
\label{APPA}
In this appendix we are going to recap the essentials of the fully inhomogeneous 
description of pre-decoupling plasmas already introduced in Eqs. (\ref{NH1})--(\ref{NH4}). We 
will follow here the formalism developed in Ref. \cite{mg2} and describe 
the fully inhomogeneous geometry in terms  of the ADM decomposition \cite{ADM1,ADM2}:
\begin{eqnarray}
&& g_{00} = N^2 - N_{k} N^{k},\qquad g_{ij} = - \gamma_{ij},\qquad g_{0i} = - N_{i},
\nonumber\\
&& g^{00} = \frac{1}{N^2},\qquad g^{ij} = \frac{N^{i} \, N^{j}}{N^2}- \gamma^{ij},\qquad 
g^{0i} = - \frac{N^{i}}{N^2}.
\label{ADM1}
\end{eqnarray}
In the ADM variables the extrinsic curvature 
$K_{ij}$ and the spatial components of the Ricci tensor $r_{ij}$ become:
\begin{eqnarray}
K_{ij} &=& \frac{1}{2 N} \biggl[- \partial_{\tau}\gamma_{ij} + ^{(3)}\nabla_{i}N_{j} + ^{(3)}\nabla_{j} N_{i} 
\biggr],
\label{ADM1a}\\
r_{ij} &=& \partial_{m} \, ^{(3)}\Gamma^{m}_{ij} -\partial_{j} ^{(3)}\Gamma_{i m}^{m} + ^{(3)}\Gamma_{i j}^{m} 
\,^{(3)}\Gamma_{m n}^{n} - ^{(3)}\Gamma_{j n}^{m} \,^{(3)}\Gamma_{i m}^{n}.
\label{ADM1b}
\end{eqnarray}
Defining as $T_{\mu\nu}$ as the total energy-momentum tensor of the fluid sources, the 
contracted form of the Einstein equations reads  
\begin{equation}
R_{\mu}^{\nu} = \ell_{\mathrm{P}}^2 \biggl[\biggl(T_{\mu}^{\nu} - \frac{T}{2} \delta_{\mu}^{\nu}\biggr) \biggr], \qquad T= g^{\mu\nu} T_{\mu\nu} = T_{\mu}^{\mu}.
\label{FORM1}
\end{equation}
As in the bulk of the paper we are now going to focus 
on the situation where the shift vectors vanish and the lapse function 
is homogeneous but time dependent (i.e. $N(\vec{x}, \tau) = N(\tau)$). 
The $(0\,0)$, $(i\,j)$ and $(0\,i)$ components of Eq. (\ref{FORM1}) become then:
\begin{eqnarray}
&& \partial_{\tau} K - N \mathrm{Tr} K^2 + \nabla^2 N = N \ell_{\mathrm{P}}^2 \biggl\{ \frac{3 p + \rho}{2} 
 + ( p + \rho) \, u^2 \biggr\},
\label{00}\\
&& \nabla_{i} K - \nabla_{k} K^{k}_{i} = N \ell_{\mathrm{P}}^2 u^{0} \,u_{i} ( p + \rho),
\label{0i}\\
&& \partial_{\tau} K_{i}^{j} - N K K_{i}^{j} - N r_{i}^{j} + \nabla_{i} \nabla^{j} N = \ell_{\mathrm{P}}^2 N\biggl[ \frac{p - \rho}{2} \delta_{i}^{j} - ( p + \rho) u_{i} u^{j}],
\label{ij}
\end{eqnarray}
where $u^2 = u^{i} \, u^{j} \gamma_{i j}$. The electron and ion velocities appearing in Eqs. (\ref{NH1}) and (\ref{NH2}) 
reduce, in the conformally flat case (i.e. $N(\tau) \to a(\tau)$ and $\gamma_{ij}(\vec{x}, \tau) \to 
a^2(\tau) \delta_{ij}$) to the velocity fields appearing in Eqs. (\ref{S2}), (\ref{S6}) and (\ref{S7}). 
In the fully inhomogeneous case, 
the evolution equations for the velocities of the electrons, ions and photons can be written, respectively, as 
\begin{eqnarray}
\partial_{\tau} v_{\mathrm{e}}^{k} + N \partial^{k} N - {\mathcal G}^{k}_{j} v_{\mathrm{e}}^{j} &=& - \frac{e \tilde{n}_{\mathrm{e}} N^2}{ \rho_{\mathrm{e}} \sqrt{\gamma}} \biggl[ E^{k} + (\vec{v}_{\mathrm{e}} \times \vec{B})^{k} \biggr] 
\nonumber\\
&+& N \Gamma_{\mathrm{ei}} (v_{\mathrm{i}}^{k} - v_{\mathrm{e}}^{k}) + \frac{4}{3} 
\frac{\rho_{\gamma}}{\rho_{\mathrm{e}}} N \Gamma_{\mathrm{e} \gamma}(v_{\gamma}^{k} - v_{\mathrm{e}}^{k}),
\label{Av1}\\
\partial_{\tau} v_{\mathrm{i}}^{k} + N \partial^{k} N - {\mathcal G}^{k}_{j} v_{\mathrm{i}}^{j} &=& \frac{e \tilde{n}_{\mathrm{i}} N^2}{ \rho_{\mathrm{i}} \sqrt{\gamma}} \biggl[ E^{k} + (\vec{v}_{\mathrm{i}} \times \vec{B})^{k} \biggr] 
\nonumber\\
&+& N\, \frac{\rho_{\mathrm{e}}}{\rho_{\mathrm{i}}}\, \Gamma_{\mathrm{ie}} (v_{\mathrm{e}}^{k} - v_{\mathrm{i}}^{k}) + \frac{4}{3} \frac{\rho_{\gamma}}{\rho_{\mathrm{i}}} N \Gamma_{\mathrm{i} \gamma}(v_{\gamma}^{k} - v_{\mathrm{i}}^{k}),
\label{Av2}\\
\partial_{\tau} v_{\gamma}^{k} + N \partial^{k} N- \biggl[ {\mathcal G}_{j}^{k} - \frac{N K}{3} \delta_{j}^{k}\biggr] v_{\gamma}^{j} &=& - \frac{N^2}{4 \rho_{\gamma}} \partial_{m} \biggl(\rho_{\gamma} \gamma^{m k} \biggr)  
\nonumber\\
&+& 
N \Gamma_{\gamma \mathrm{e}} ( v_{\mathrm{e}}^{k} - v_{\gamma}^{k}) + 
N \Gamma_{\gamma \mathrm{i}} ( v_{\mathrm{i}}^{k} - v_{\gamma}^{k}),
\label{Av2a}
\end{eqnarray}
where 
\begin{equation}
 {\mathcal G}^{k}_{j}= \biggl[\frac{\partial_{\tau} N}{N} \delta_{j}^{k} + 2 N K_{j}^{k}\biggr].
\label{Av3}
\end{equation} 
As in the conformally flat case the evolution equations of the electrons and of the ions 
can be combined by defining the center of mass velocity of the electron-ion system 
$v_{\mathrm{b}}^{k} = (m_{\mathrm{e}} v_{\mathrm{e}}^{k} + m_{\mathrm{i}} v_{\mathrm{i}}^{k})/(m_{\mathrm{e}} + m_{\mathrm{i}})$ so that the the effective evolution equations for the baryon-lepton-photon fluid become
\begin{eqnarray}
\partial_{\tau} \rho_{\gamma} &=& \frac{4}{3} K N \rho_{\gamma} - \frac{4}{3} N \partial_{k}\biggl( \frac{\rho_{\gamma}}{N}\,v_{\gamma}^{k}\biggr),
\label{Av4}\\
\partial_{\tau} v_{\mathrm{b}}^{k} &=& {\mathcal G}_{j}^{k} v_{\mathrm{b}}^{j}  - N \partial^{k} N  + \frac{(\vec{J} \times \vec{B})^{k} N^2}{ \gamma\, \rho_{\mathrm{b}} ( 1 + m_{\mathrm{e}}/m_{\mathrm{i}})} + 
\frac{4}{3} \frac{\rho_{\gamma}}{\rho_{\mathrm{b}}} N \Gamma_{\gamma\mathrm{e}} (v_{\gamma}^{k} - v_{\mathrm{b}}^{k}), 
\label{Av5}\\
\partial_{\tau} v_{\gamma}^{k} &=& \biggl[ {\mathcal G}_{j}^{k} - \frac{N K}{3} \delta_{j}^{k}\biggr] v_{\gamma}^{j} - \frac{N^2}{4 \rho_{\gamma}} \partial_{m} \biggl(\rho_{\gamma} \gamma^{m k} \biggr) - N \partial^{k} N +
N \Gamma_{\gamma \mathrm{e}} ( v_{\mathrm{b}}^{k} - v_{\gamma}^{k}),
\label{Av6}
\end{eqnarray}
where $v_{\gamma}^{k}$ and $\rho_{\gamma}$ denote, respectively, the photon velocity and the photon 
energy density. 
\renewcommand{\theequation}{B.\arabic{equation}}
\setcounter{equation}{0}
\section{Some relevant correlators}
\label{APPB}
The correlator appearing in Eq. (\ref{F12}), i.e.
\begin{equation}
{\mathcal T}_{b m b' m'} =  \langle \biggl[ h_{a b} \, h_{ a q} \, h_{q m}\biggr]_{\vec{x}} \, \biggl[ h_{a'\, b'} h_{a' \,q'} h_{q' m'}\biggr]_{\vec{y}} \rangle,
\label{AB1}
\end{equation}
must be computed in terms of the corresponding two-point functions in real space (see Eq. 
(\ref{F6})).  The general form of the two-point function in real space has been already mentioned in Eq. (\ref{F9A}) 
and the functions $G_{i}(r)$ (with $i = 1,\,...\, 5$) are given by:
\begin{eqnarray}
G_{1}(r) &=& F_{1}(r) + \frac{2}{r} \frac{\partial F_{2}}{\partial r}+ \frac{1}{r} \frac{\partial}{\partial r} \biggl( \frac{1}{r} \frac{\partial F_{3}}{\partial r}\biggr),
\label{F9B}\\
G_{2}(r) &=& \frac{1}{r} \frac{\partial}{\partial r} \biggl( \frac{1}{r} \frac{\partial F_{3}}{\partial r}\biggr) - F_{1}(r) - \frac{2}{r} \frac{\partial F_{2}}{\partial r},
\label{F9C}\\
G_{3}(r) &=& \frac{1}{r} \frac{\partial}{\partial r}\biggl[ \frac{1}{r} \frac{\partial}{\partial r} \biggl( \frac{1}{r} \frac{\partial F_{3}}{\partial r}\biggr)\biggr] 
- \frac{1}{r} \frac{\partial}{\partial r} \biggl( \frac{1}{r} \frac{\partial F_{2}}{\partial r}\biggr),
\label{F9D}\\
G_{4}(r) &=& \frac{1}{r} \frac{\partial}{\partial r}\biggl[ \frac{1}{r} \frac{\partial}{\partial r} \biggl( \frac{1}{r} \frac{\partial F_{3}}{\partial r}\biggr)\biggr] 
+ \frac{1}{r} \frac{\partial}{\partial r} \biggl( \frac{1}{r} \frac{\partial F_{2}}{\partial r}\biggr),
\label{F9E}\\
G_{5}(r) &=& \frac{1}{r} \frac{\partial}{\partial r}\biggl\{ \frac{1}{r} \frac{\partial}{\partial r}\biggl[ \frac{1}{r} \frac{\partial}{\partial r}
\biggl( \frac{1}{r}\frac{\partial F_{3}}{\partial r} \biggr)\biggr]\biggr\},
\label{F9F}
\end{eqnarray}
where $F_{1}(r)$, $F_{2}(r)$ and $F_{3}(r)$ are fully determined once the power spectrum is known and are 
defined as:
\begin{eqnarray}
F_{1}(r) &=& \frac{1}{4} \int \, \frac{d k}{k} \, {\mathcal P}_{\mathrm{T}}(k,\tau) \frac{\sin{k r}}{k r}, \qquad F_{2}(r)=
 \frac{1}{4} \int \, \frac{d k}{k^3} \, {\mathcal P}_{\mathrm{T}}(k,\tau) \frac{\sin{k r}}{k r},
 \nonumber\\
F_{3}(r) &=& \frac{1}{4} \int \, \frac{d k}{k^5} \, {\mathcal P}_{\mathrm{T}}(k,\tau) \frac{\sin{k r}}{k r}.
\label{F9G}
\end{eqnarray}
Using Eq. (\ref{F9G}) inside Eqs. (\ref{F9B})--(\ref{F9F}) we have that 
\begin{eqnarray}
G_{1}(r) &=& \frac{1}{4} \int \frac{d k}{k} \, {\mathcal P}_{\mathrm{T}}(k,\tau) \biggl[ \biggl( 1 - \frac{1}{k^2 r^2}\biggr) j_{0}(k r) + \biggl( \frac{3}{k^2 r^2} - 2\biggr)\frac{j_{1}(k r)}{k r}
\biggr],
\label{F10B}\\
G_{2}(r) &=& \frac{1}{4} \int \frac{d k}{k} \, {\mathcal P}_{\mathrm{T}}(k,\tau) \biggl[\biggl( 2 + \frac{3}{k^2 r^2}\biggr) \frac{j_{1}(k r)}{k r} - \biggl(1 + \frac{1}{k^2 r^2}\biggr) j_{0}(k r) 
\biggr],
\label{F10C}\\
G_{3}(r) &=& \frac{1}{4} \int  k\, d k {\mathcal P}_{\mathrm{T}}(k,\tau) \biggl[ \frac{j_{0}(k r)}{k^2 r^2} \biggl( 1 + \frac{5}{k^2 r^2} \biggr) - 
 \frac{j_{1}(k r)}{k^3 r^3}\biggl( 2 + \frac{15}{k^2 r^2}\biggr)\biggr],
\label{F10D}\\
G_{4}(r) &=& \frac{1}{4} \int  k\, d k{\mathcal P}_{\mathrm{T}}(k,\tau) \biggl[ \frac{j_{0}(k r)}{k^2 r^2} \biggl( -1 + \frac{5}{k^2 r^2}\biggr) + \frac{j_{1}(k r)}{k^3 r^3}\biggl( 4 - \frac{15}{k^2 r^2}\biggr)\biggr],
\label{F10E}\\
G_{5}(r) &=& \frac{1}{4} \int k^3 d k {\mathcal P}_{\mathrm{T}}(k,\tau)   \biggl[ \frac{j_{0}(k r)}{k^4 r^4}\biggl( 1 -\frac{35}{k^2 r^2} \biggr) 
- \frac{5 j_{1}(k r)}{k^5 r^5} \biggl( 2 - \frac{21}{k^2 r^2}\biggr)\biggr],
\label{F10F}
\end{eqnarray}
where $j_{0}(k r)$ and $j_{1}(k r)$ are spherical Bessel functions of zeroth and first-order
\cite{abr1,abr2}
\begin{equation}
j_{0}(k r) = \frac{\sin{k r}}{k r}, \qquad j_{1}(k r) = \frac{\sin{k r}}{k^2 r^2} - \frac{\cos{k r}}{k r}.
\label{BS}
\end{equation}
It is useful to compare the two different asymptotic limits of the various $G_{i}(r)$, i.e. for $k r < 1$ and for $k r > 1$. 
In the limit $k r > 1$ we have that: 
\begin{eqnarray}
G_{1}(r) &\to& \frac{1}{4} \int \frac{d k}{k} \, {\mathcal P}_{\mathrm{T}}(k,\tau) j_{0}(k r), \qquad G_{2}(r) \to  - G_{1}(r),
\label{LIM1}\\
G_{3}(r) &\to& \frac{1}{4} \int k\, d k\, {\mathcal P}_{\mathrm{T}}(k,\tau) \frac{j_{0}(k r)}{k^2 r^2}, \qquad G_{4}(r) \to - G_{3}(r),
\label{LIM2}\\
G_{5}(r) &\to&  \frac{1}{4} \int k^{3} d k \, {\mathcal P}_{\mathrm{T}}(k,\tau) \frac{j_{0}(k r)}{k^4 r^4}.
\label{LIM3}
\end{eqnarray}
Conversely, in the limit $k r < 1$, Eqs. (\ref{F10B})--(\ref{F10F}) imply:
\begin{eqnarray}
 G_{1}(r) &\to& \frac{1}{10} \int \frac{d k}{k} {\mathcal P}_{\mathrm{T}}(k,\tau)\biggl[ 1 - \frac{11}{42} k^2 r^2\biggr],
\label{LIM4}\\
G_{2}(r) &\to& - \frac{1}{15}  \int \frac{d k}{k} {\mathcal P}_{\mathrm{T}}(k,\tau)\biggl[ 1 - \frac{5 k^2 r^2}{14}\biggr],
\label{LIM5}\\
G_{3}(r) &\to& - \frac{2}{105} \int  k d k {\mathcal P}_{\mathrm{T}}(k,\tau)\biggl[ 1 - \frac{5}{72} k^2 r^2\biggr],
\label{LIM6}\\
G_{4}(r) &\to& \frac{1}{70} \int k d k {\mathcal P}_{\mathrm{T}}(k,\tau)\biggl[ 1 - \frac{2\,  k^2 r^2}{27}\biggr],
\label{LIM7}\\
G_{5}(r) &\to& \frac{1}{3780} \int k^3\,d k  {\mathcal P}_{\mathrm{T}}(k,\tau)\biggl[ 1 - \frac{k^2 r^2}{22}\biggr].
\label{LIM8}
\end{eqnarray}
The explicit form of the two-point function $G_{i j m n}$ implies that the six-point function 
appearing in the correlator of the vorticity can be expressed as 
\begin{equation}
{\mathcal T}_{b m b' m'}  = \sum_{\nu=1}^{5} {\mathcal T}^{(\nu)}_{b\, m\, b'\, m'},
\label{AB2}
\end{equation}
where the $5$ distinct contributions correspond to
\begin{eqnarray}
{\mathcal T}^{(1)}_{b\, m\, b'\, m'} &=& \overline{G}_{a b a q} \biggl[ G_{q m a' b'} \overline{G}_{a' q' q' m'}
+ G_{q m a' q'} \overline{G}_{b' a' q' m'} + G_{q m q' m'} \overline{G}_{a' q' a' b'}\biggr],
\label{AB3}\\
{\mathcal T}^{(2)}_{b\, m\, b'\, m'} &=& \overline{G}_{a b q m} \biggl[ G_{a q a' b'} \overline{G}_{a' q' q' m'}
+ G_{a q a' q'} \overline{G}_{a' b' q' m'} + G_{a q q' m'} \overline{G}_{a' b' a' q'}\biggr],
\label{AB4}\\
{\mathcal T}^{(3)}_{b\, m\, b'\, m'} &=& G_{a b a' b'} \overline{G}_{a q q m} \overline{G}_{a' q' q' m'}
\nonumber\\
&+& G_{a b a' b'} \biggl( G_{a q a' q'} G_{q m q' m'} + G_{a q q' m'} G_{q m a' q'}\biggr),
\label{AB5}\\
{\mathcal T}^{(4)}_{b\, m\, b'\, m'} &=& G_{a b q' a'} \overline{G}_{a q q m} \overline{G}_{a' b' q' m'}
\nonumber\\
&+&  G_{a b q' a'}\biggl( G_{a q a' b'} G_{q m q' m'} + G_{a q q' m'} G_{q m a' b'}\biggr),
\label{AB6}\\
{\mathcal T}^{(5)}_{b\, m\, b'\, m'} &=&  G_{a b q' m'} \overline{G}_{a q q m} \overline{G}_{a' b' a' q'}
\nonumber\\
&+& G_{a b q' m'} \biggl( G_{a q a' b'} G_{q m a' q'} + G_{a q a' q'} G_{q m a' b'}\biggr).
\label{AB7}
\end{eqnarray}
The overline signifies that the corresponding correlator is evaluated in the limit $r\to 0$. According to 
Eqs. (\ref{LIM4})--(\ref{LIM8}) this limit is non-singular. 
Notice finally that in terms of ${\mathcal T}_{b m b' m'}$ the correlation function 
of the magnetic field can also be written, with shorthand notation, as 
\begin{eqnarray}
\langle B^2(r) \rangle &=& {\mathcal J}(\tau,w)  \epsilon^{j m i}\, \epsilon^{j' m' i} \frac{\partial^2}{\partial y^{j'}\, \partial y^{b'}} \frac{\partial^2 }{\partial x^{j} \, \partial x^{b}} {\mathcal T}_{b m b' m'}
\nonumber\\
{\mathcal J}(\tau, w) &=& \frac{m_{\mathrm{i}}^2}{\alpha_{\mathrm{em}}} \frac{{\mathcal H}^2\, a_{1}^2}{ 9 {\mathcal H}_{1}^4 (w+ 1)^2} \biggl(\frac{a}{a_{1}}\biggr)^{6 w + 4}.
\label{short}
\end{eqnarray}
The real space 
approach is more effective and convenient for an explicit estimate of the vorticity
and the idea is therefore to express the correlation functions in real space, take the 
appropriate derivatives and then expand the result in the desired limit. 
Denoting with $R = r/r_{\mathrm{p}}$ and with $x = k/k_{\mathrm{p}}$, 
the integrals over $k$ appearing in Eqs. (\ref{F10B})--(\ref{F10F}) 
can be computed explicitly by changing variable and by using the following 
pair of relations \cite{grad}:
\begin{eqnarray}
 \int_{1}^{\infty} x^{n - m} \sin{x R} \, d x&=& \frac{R}{m - n -2} \,\,  _{1}F_{2}\biggl[ a_{1}; b_{1}, b_{2}; - \frac{R^2}{4}  \biggr] 
\nonumber\\
&+& R^{m - n -1} \cos{\biggl[\frac{\pi(m - n)}{2}\biggr]} 
\Gamma[1 - m + n],
\label{AB8a}\\
 \int_{1}^{\infty} x^{n - m} \cos{x R}\, d x &=& \frac{1}{m - n -2} \,\,  _{1}F_{2}\biggl[ \tilde{a}_{1}; \tilde{b}_{1}, \tilde{b}_{2}; - \frac{R^2}{4}  \biggr] 
\nonumber\\
&+& R^{m - n -1} \sin{\biggl[\frac{\pi(m - n)}{2}\biggr]} \Gamma[1 - m + n],
\label{AB8b}
\end{eqnarray}
where $n < m$ (i.e. $(n - m)$ is negative).
In Eqs. (\ref{AB8a}) and (\ref{AB8b})  $   _{p}F_{q}[ a_{1}....a_{p}; b_{1}, ... b_{q}; z] $ 
denotes  the generalized hypergeometric function of argument $z$; in the 
case of Eqs. (\ref{AB8a}) and (\ref{AB8b}) , $p= 1$, $q=2$ and 
\begin{eqnarray}
&&a_{1} = 1 + \frac{n - m}{2},\qquad b_{1} = \frac{3}{2},\qquad b_{2} = 2 + \frac{n - m}{2},
\label{AB9a}\\
&& \tilde{a}_{1} = a_{1} - \frac{1}{2},\qquad \tilde{b}_{1} = b_{1} -1,\qquad \tilde{b}_{2} = b_{2} - \frac{1}{2}.
\label{AB9b}
\end{eqnarray}
The integrals are taken from $1$ to infinity since the integral over $k$ starts from $k_{\mathrm{p}}$ implying that the lower limit of integration in $x$ is 1. Equations (\ref{AB8a}) and (\ref{AB8b}) 
can be used to derive the real space form of the correlator of Eq. (\ref{F6}).  Using Eqs.  
(\ref{AB8a}) and (\ref{AB8b}), the explicit form of Eqs. (\ref{F10B})--(\ref{F10F}) can be derived and inserted 
into Eq. (\ref{F9A}) whose explicit form determines the real-space expression of the two-point functions 
of the vorticities through Eqs. (\ref{AB3})--(\ref{AB7}). After taking the appropriate derivatives 
the obtained result can be expanded in the wanted limits (e. g. $R\gg 1$ or $R\ll 1$). 
The explicit real-space expressions of Eqs. (\ref{F10B})--(\ref{F10F}) are typically rather lengthy  but they are 
conceptually straightforward. This is why they will be omitted and only an example will be given.
Even if the scales interesting for section \ref{sec6} will be the ones close to the galactic scale, consider, for instance the expression of $G_{1}(R)$ in the opposite limit, i.e.  comoving scales much larger
than $r_{\mathrm{eq}}$. In this case the expressions simplify since ${\mathcal M}(k, k_{\mathrm{eq}}, \tau) \to 1$. 
Therefore, using Eqs. (\ref{AB8a}) and (\ref{AB8b}), Eq. (\ref{F10B}) becomes:
\begin{eqnarray}
G_{1}(R) &=& \frac{{\mathcal A}_{\mathrm{T}}}{4}\biggl\{  \frac{1}{n_{\mathrm{T}}} \, _{1}F_{2}\biggl[
\frac{n_{\mathrm{T}}}{2};\, \frac{3}{2}, \frac{n_{\mathrm{T}}}{2}; - \frac{R^2}{4}\biggr]
\nonumber\\
&+& \frac{3}{4(n_{\mathrm{T}} -4) R^4}\biggl[\, _{1}F_{2}\biggl[ -2 + \frac{n_{\mathrm{T}}}{2}; \frac{1}{2}, -1 + 
\frac{n_{\mathrm{T}}}{2}; - \frac{R^2}{4}\biggr]  
\nonumber\\
&-& 
\, _{1} F_{2}\biggl[ -2 + \frac{n_{\mathrm{T}}}{2}; \frac{3}{2}, -1 + 
\frac{n_{\mathrm{T}}}{2}; - \frac{R^2}{4}\biggr]\,\biggr]
\nonumber\\
&+& \frac{1}{4(n_{\mathrm{T}} -2) R^2}\biggl[\,  3\, _{1}F_{2}\biggl[ -1 + \frac{n_{\mathrm{T}}}{2}; \frac{3}{2},  
\frac{n_{\mathrm{T}}}{2}; - \frac{R^2}{4}\biggr] 
\nonumber\\
&-& 2\, _{1}F_{2}\biggl[ -1 + \frac{n_{\mathrm{T}}}{2}; \frac{1}{2},  
\frac{n_{\mathrm{T}}}{2}; - \frac{R^2}{4}\biggr] \, \biggr]
\nonumber\\
&-& \frac{1}{4 R^{n_{\mathrm{T}}}} \cos{\biggl(\frac{\pi n_{\mathrm{T}}}{2} \biggr)}\biggl[ \Gamma[n_{\mathrm{T}}- 4] + \Gamma[n_{\mathrm{T}} -2] + 3 \Gamma[n_{\mathrm{T}} - 5] 
\nonumber\\
&+& 3 \Gamma[n_{\mathrm{T}} - 3] + 
\Gamma[n_{\mathrm{T}}-1]\biggr]\, \biggr\}.
\label{EXG}
\end{eqnarray}

To conclude this appendix let us show that 
the expression of $G_{i j m n}$ given in Eq. (\ref{F6}) coincides with the result directly 
obtainable in the particular case where the tensor modes of the geometry are quantized in terms of gravitons. When $h_{i j}(\vec{x},\tau)$ is a field operator  its expression 
can be written as \cite{mgt1,mgt2}
 \begin{equation}
 \hat{h}_{ij}(\vec{x},\tau)= \frac{\sqrt{2} \ell_{\mathrm{P}}}{(2\pi)^{3/2} a(\tau)} \sum_{\lambda}
  \int d^{3}k\,\, \epsilon_{ij}^{(\lambda)}(\hat{k})\, 
 \biggl[ \hat{a}_{\vec{k},\lambda} \,f_{k,\lambda}(\tau) e^{- i \vec{k} \cdot \vec{x}} +  \hat{a}_{\vec{k},\lambda}^{\dagger}\, f_{k,\lambda}^{*}(\tau) 
 e^{ i \vec{k} \cdot \vec{x}}\biggr].
 \label{F10}
 \end{equation}
which also implies, using the properties of the creation and annihilation operators, 
\begin{equation}
 G_{i j m n}(r)= \langle \hat{h}_{i j}(\vec{x},\tau) \, \hat{h}_{m n}(\vec{y},\tau) \rangle = 
 \int \frac{d k }{k} {\mathcal P}_{\mathrm{T}}(k, \tau) \, {\mathcal Q}_{i j m n}(\hat{k}) \, j_{0}(k r),
  \label{F11A}
 \end{equation}
  where 
 \begin{eqnarray}
 {\mathcal P}_{\mathrm{T}}(k,\tau) &=& 4\ell_{\mathrm{P}}^2\frac{k^3}{\pi^2 a^2(\tau)} |f_{k}(\tau)|^2,
 \label{F11AA}\\
 {\mathcal Q}_{i j m n} &=& \frac{1}{4} \sum_{\lambda} \epsilon_{ij}^{(\lambda)} \, \epsilon_{m n}^{(\lambda)} = 
 \frac{1}{4} \biggl[P_{m i}(\hat{k}) P_{n j}(\hat{k}) + P_{m j}(\hat{k}) P_{n i}(\hat{k}) - P_{i j}(\hat{k}) P_{m n}(\hat{k}) \biggr];
\label{F11B} 
\end{eqnarray}
$P_{ij}(\hat{k}) = (\delta_{i j} - \hat{k}_{i} \hat{k}_{j})$ with $\hat{k}^{i} = k^{i}/|\vec{k}|$. In Eq. (\ref{F11A}) it has been used that 
\begin{equation}
\langle 0| \, \hat{a}_{\vec{p},\mu}\,  \hat{a}^{\dagger}_{\vec{p},\,\lambda}|0 \rangle = \delta^{(3)}(\vec{k} - \vec{p}) \delta_{\lambda\mu}.
\label{Q1}
\end{equation}
Furthermore, to derive Eq. (\ref{F11B}), the explicit form of the two tensor polarizations can be written, in explicit terms, as 
\begin{equation}
\epsilon_{ij}^{(\oplus)}(\hat{k}) = (\hat{a}_{i} \hat{a}_{j} - \hat{b}_{i} \hat{b}_{j}), \qquad 
  \epsilon_{ij}^{(\otimes)}(\hat{k}) = (\hat{a}_{i} \hat{b}_{j} + \hat{b}_{i} \hat{a}_{j}),
\label{F9}
\end{equation}
where $\hat{a}$, $\hat{b}$ and $\hat{k}$ are three mutually orthogonal unit vectors. 
\end{appendix}

\end{document}